\documentclass[showpacs,preprintnumbers,amsmath,amssymb]{revtex4}
\usepackage{amsmath,amssymb,graphics,epsfig,subfigure}
\usepackage{color}

\begin{document}
\renewcommand{\baselinestretch}{1.3}

\title{QED effects on phase transition and Ruppeiner geometry of Euler-Heisenberg-AdS black holes}

\author{Xu Ye, Zi-Qing Chen, Ming-Da Li, Shao-Wen Wei \footnote{Corresponding author. E-mail: weishw@lzu.edu.cn}}

\affiliation{$^{1}$Lanzhou Center for Theoretical Physics, Key Laboratory of Theoretical Physics of Gansu Province, School of Physical Science and Technology, Lanzhou University, Lanzhou 730000, People's Republic of China,\\
 $^{2}$Institute of Theoretical Physics $\&$ Research Center of Gravitation,
Lanzhou University, Lanzhou 730000, People's Republic of China,\\
 $^{3}$Academy of Plateau Science and Sustainability, Qinghai Normal University, Xining 810016, P. R. China}

\begin{abstract}
Taking the quantum electrodynamics (QED) effect into account, we study the black hole phase transition and Ruppeiner geometry for the Euler-Heisenberg anti-de Sitter black hole in the extended phase space. For negative and small positive QED parameter, we observe a small/large black hole phase transition and reentrant phase transition, respectively. While a large positive value of the QED parameter ruins the phase transition. The phase diagrams for each case are explicitly exhibited. Then we construct the Ruppeiner geometry in the thermodynamic parameter space. Different features of the corresponding scalar curvature are shown for both the small/large black hole phase transition and reentrant phase transition cases. Of particular interest is that an additional region of positive scalar curvature indicating dominated repulsive interaction among black hole microstructure is present for the black hole with a small positive QED parameter. Furthermore, the universal critical phenomena are also observed for the scalar curvature of the Ruppeiner geometry. These results indicate that the QED parameter has a crucial influence on the black hole phase transition and microstructure.
\end{abstract}


\maketitle


\section{ Introduction}\label{a}

Due to the earliest pioneering work of Hawking and Bekenstein on the temperature and entropy of black holes, it has shown that there is a deep connection between gravity, quantum mechanics, and thermodynamics \cite{Bekensteing73,hawking1975particle,hawking1976black}. Equipped with the established four laws of black hole thermodynamics \cite{bardeen1973four}, the study of the thermodynamics becomes one of the increasingly active areas in black hole physics. Motivated by the anti-de Sitter/conformal field theory (AdS/CFT) correspondence \cite{witten1998anti,maldacena1998large}, the Hawking-Page phase transition \cite{hawking1983thermodynamics} between the stable large Schwarzschild AdS black hole and thermal space was interpreted as the confinement/deconfinement phase transition of the gauge field \cite{witten1998thermal}. The phase transition was also extended to the charged and rotating AdS black hole cases \cite{chamblin1999charged,chamblin1999holography,caldarelli2000thermodynamics}.

Recently, it was found in the AdS space that, by interpreting the cosmological constant as thermodynamic pressure \cite{kastor2009enthalpy,dolan2011pressure,cvetivc2011black}, the black hole systems are analog to everyday thermodynamical systems. In this extended thermodynamic phase space, the charged small/large AdS black hole phase transition is similar to the gas-liquid phase transition of Van der Waals (VdW) fluid \cite{Kubiznak:2012wp}. Subsequently, more interesting black hole phase transitions and phase structures, such as reentrant phase transition \cite{NaveenaKumara:2020biu,altamirano2013reentrant}, isolated critical point \cite{dolan2014isolated}, triple point \cite{altamirano2014kerr,wei2014triple,frassino2014multiple}, and superfluid black hole phase \cite{hennigar2017superfluid}, have been uncovered.

Understanding black hole microstructure has to be a huge challenge. Although the string theory \cite{Strominger:1996sh,Maldacena:1996gb,Callan:1996dv,Horowitz:1996fn}, fuzzy ball model \cite{Lunin:2001jy,Lunin:2002qf}, and pierced horizons \cite{Rovelli:1996dv} have made great progress, more questions remain to be solved. As proposed in Refs. \cite{Wei:2015iwa,Wei:2019uqg}, the study of black hole phase transition can also be applied to this challenge with the assumption that the micro-degree of freedom is measured by the underlying molecules of the black hole. Combining with the Ruppeiner geometry \cite{ruppeiner1995riemannian}, the characteristic black hole microstructure was tested. Its scalar curvature of the corresponding geometry is an important tool for exploring the microstructure of a black hole. Via the empirical observation that the positive or negative scalar curvature corresponds to the repulsive or attractive interaction among these underlying black hole molecules. Such empirical result was supported by many studies of different fluid systems, such as ideal fluids \cite{ingarden1978information}, Van der Waals (VdW) fluids \cite{janyszek1990riemannian}, one-dimensional Ising models, quantum gases \cite{ruppeiner1981application,janyszek1989magnetic}. More significantly, when considering the microscopic model and the equation of state, a possible interpretation of the empirical observation was given and the corresponding molecular potential is constructed in Ref. \cite{Wei:2021hva}.

In view of the "hard-core" model, there is only the dominated attractive interaction for the VdW fluid. However for the charged AdS black holes, the dominated repulsive interaction emerges for the small black hole with high temperature \cite{Wei:2019yvs}, which uncovers an interesting phenomenon for the black hole microstructure. After generalizing the study to the modified gravity \cite{Wei:2019ctz}, it was found that the dominated repulsive interaction may not emerge, while the attractive interaction is universal.

On the other hand, black hole solutions with nonlinear electrodynamics gain great interest. In particular, this theory can be viewed as a low-energy limit from string theory or D-brane physics, where Abelian and non-Abelian nonlinear electrodynamic Lagrangians could be produced. The thermodynamics of the charged Born-Infeld (BI) AdS black holes was considered in Ref. \cite{Gunasekaran:2012dq}. The present of the BI vacuum polarization governs a rich black hole phase transition. The study was also extended to higher dimensions, and the small/large black hole phase transition was found to be universal \cite{Zou:2013owa}. Another widely concerned nonperturbative one-loop effective Lagrangian of nonlinear electromagnetic fields originates from Heisenberg and Euler \cite{heisenberg1936folgerungen}. Within the QED framework, it was reformulated by Schwinger \cite{schwinger1951gauge}. The black hole solutions corresponding to the effective Lagrangian have been worked out \cite{ruffini2013einstein}. In Refs. \cite{Magos:2020ykt,Li:2021ygi}, the first law of black hole thermodynamics and Smarr formula were found to be consistent with each other when the vacuum polarization parameter is included in. The phase transition was preliminarily studied. At the critical point, the standard mean field theory exponents were obtained.

Motivated by it, we in this paper will thoroughly study the phase transition for the charged Euler-Heisenberg (EH)-AdS black holes by taking the QED effect into account. The small/large black hole phase transition and reentrant phase transition are found to exist in different regions of the parameter space. The phase diagrams and phase structures are completely exhibited. Then based on its phase diagram, we constructed the Ruppeiner geometry for the charged EH-AdS black hole. The feature of the scalar curvature is also obtained. Further employing with the empirical observation of the Ruppeiner geometry, we disclose the particular interesting property of the microstructure for the black holes.

This article is organized as follows. In Sec. \ref{b}, we firstly review the thermodynamics for the charged EH-AdS black hole. Then in different regions of the parameter space, we study the small/large black hole phase transition and reentrant phase transition, respectively. The phase diagrams are clearly exhibited. In Sec. \ref{c}, the Ruppeiner geometry is constructed. Employing with the corresponding scalar curvature, we explore the characteristic black hole microstructure under the small/large black hole phase transition and reentrant phase transition. Novel properties are found. Finally, the conclusions and discussions are given in Sec. \ref{d}.

\section{Thermodynamics and phase transition of Euler-Heisenberg-AdS black hole}\label{b}

In this section, we would like to briefly review the thermodynamics for the charged EH-AdS black hole with the QED correction \cite{ruffini2013einstein}, and then study its phase transition in a complete parameter space, where the small/large black hole phase transition, reentrant phase transition will be clearly exhibited.

\subsection{Thermodynamics and critical points}

The action of the EH theory with cosmological constant in four-dimensional spacetime is \cite{salazar1987duality}
\begin{equation}
S=\frac{1}{4\pi}\int_{M^4} d^4x \sqrt{-g}\left[\frac{1}{4}(R-2\Lambda)-
\mathcal{L}(F,G) \right],\label{action}
\end{equation}
\begin{equation}
\mathcal{L}(F,G)=-F+\frac{a}{2}F^2+ \frac{7a}{8} G^2. \label{Lagrangian}
\end{equation}
Here $g$, $\Lambda$, and $R$ are the determinant of the metric tensor, cosmological constant, and Ricci curvature tensor, respectively. The term $\mathcal{L}(F,G)$ is the Lagrangian of the nonlinear electrodynamics which depends on the electromagnetic invariants $F=\frac{1}{4}F_{\mu\nu}F^{\mu\nu}$ and $G=\frac{1}{4}F_{\mu \nu}{^*F^{\mu \nu}}$. The Maxwell field strength $F_{\mu \nu}=\partial_{\mu}\mathcal{A}_{\nu}-\partial_{\nu}\mathcal{A}_{\mu}$, and $\mathcal{A}_{\mu}$ is the corresponding vector potential. The parameter $a$ appearing in formula (\ref{Lagrangian}) can be used to measure the strength of the QED correction, which is related to the mass and charge of the electron, and thus we call it the QED parameter. When $a$=0, the influence of the QED term will vanish.

For a static spherically symmetric charged EH-AdS black hole solution, the line element is \cite{Magos:2020ykt}
\begin{equation}
ds^2= -f(r)dt^2 + f(r)^{-1}dr^2 + r^2(d\theta^2 +\sin^2{\theta} d\phi^2),\label{LineElement}
\end{equation}
where the metric function is given by
\begin{equation}
f(r)= 1-\frac{2M}{r}+\frac{Q^2}{r^2}-\frac{\Lambda r^2}{3}-\frac{a Q^4}{20 r^6},\label{metric}
\end{equation}
where $ M $ and $ Q $ are the mass and electric charge of the black hole, respectively. When $a=0$, this solution reduces to the Reissner-Nordstr\"{o}m (RN) AdS black hole solution.

\begin{figure}[h]
\centering
\subfigure[]{\includegraphics[width=7cm]{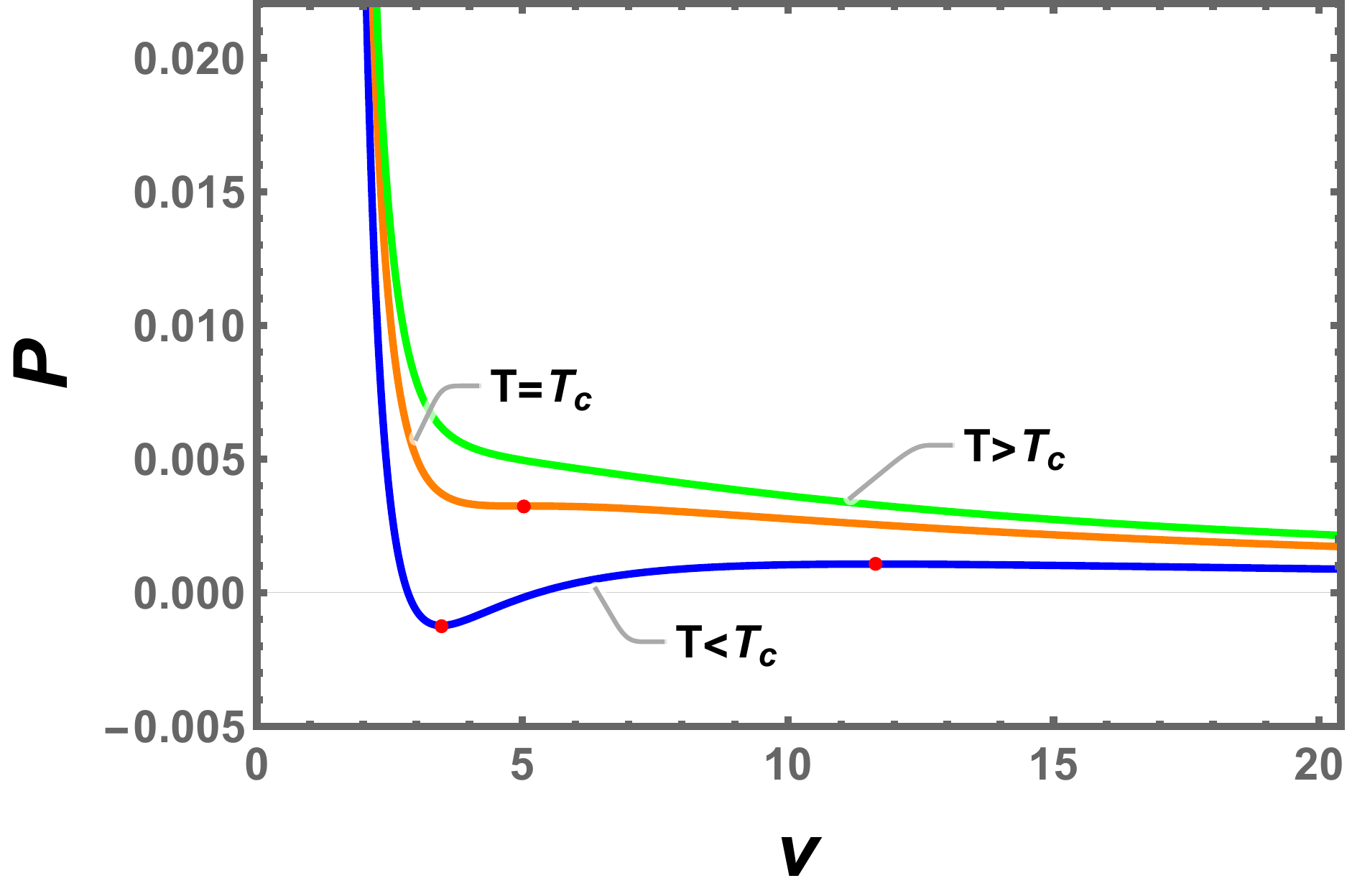} \label{fig:VdWpv_1a}}
\subfigure[]{\includegraphics[width=7cm]{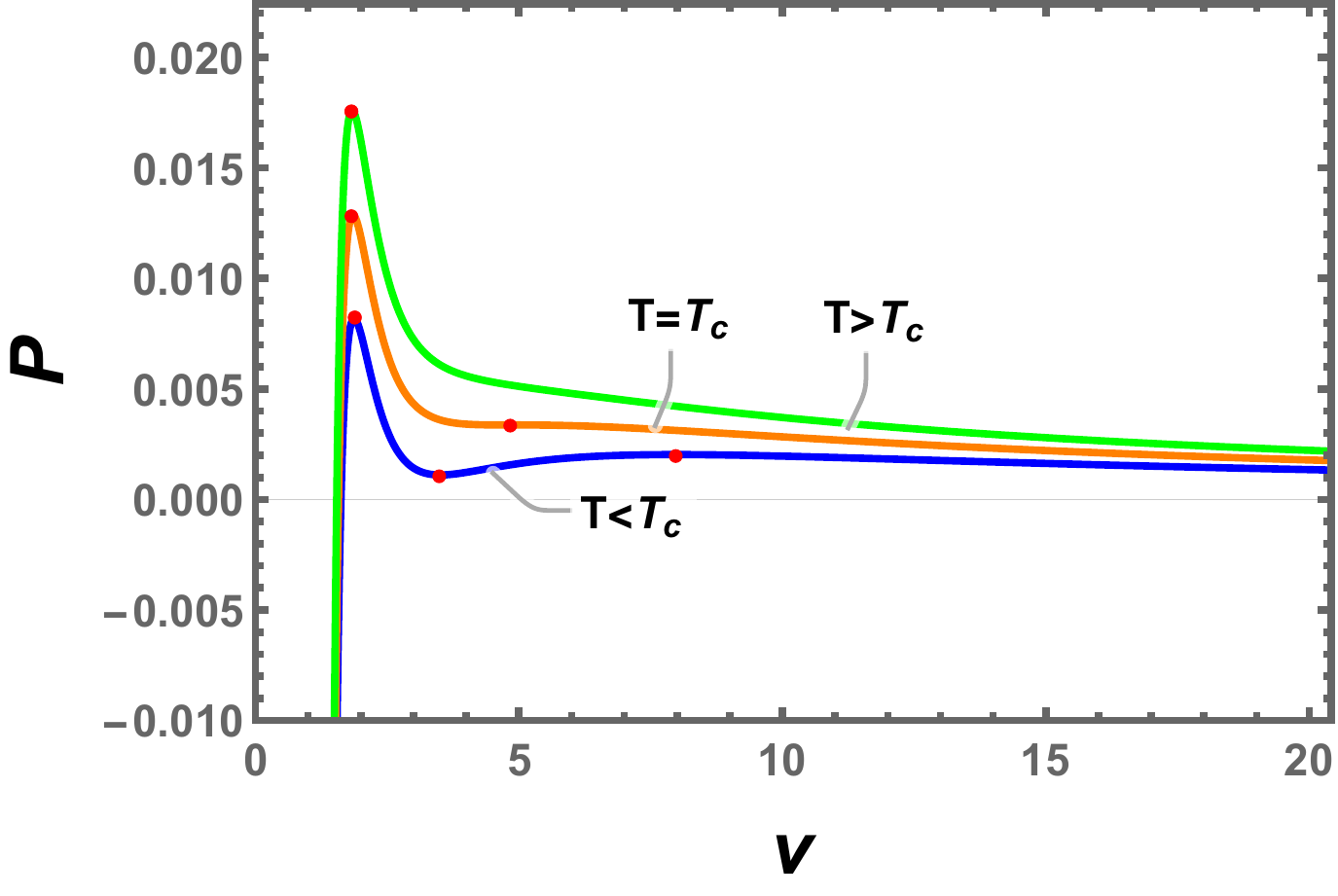} \label{fig:pv_1b}  }
\caption{The isothermal curves in $P$-$v$ plane of the charged EH-AdS black hole. (a) $a=-1.5$ ($a<0$). (b) $ a=1.0 $ ($0 \leq a \leq \frac{32}{7}Q^2 $). The temperature $T<T_c$, $T=T_c$, and $T>T_c$ from bottom to top. The extremal points are marked with red dots and we have set $Q=1$.}
\label{fig:Pvimage_1}
\end{figure}

The radius $r_+$ of the outer black hole event horizon is the largest root of $ f(r_+)=0 $, which can be obtained by solving
\begin{equation}
\frac{\Lambda}{3} r_{+}^8 - r_{+}^6-Q^2 r_{+}^{4}+2M r^{5}_{+} + \frac{Q^4 a}{20}=0 .\label{eightdeg}
\end{equation}
In term of $r_+$, the mass $M$ can be expressed as
\begin{equation}	
 M=\frac{-3 a Q^4+60 Q^2 r_+^4-20 \Lambda  r_+^8+60 r_+^6}{120	r_+^5}.
\end{equation}
Using the ``Euclidean trick", the Hawking temperature is
\begin{equation}
T=\frac{f'(r_+)}{4\pi}=\frac{1}{4\pi r_+}\left(1-\frac{Q^2}{r_+^2}+\frac{a
	Q^4}{4r_+^6}-\Lambda r_+^2\right).\label{Hawkingtem}
\end{equation}
In the extended phase space, the cosmological constants $\Lambda$ was interpreted as the pressure via $P=-\frac{\Lambda}{8\pi}$ \cite{kastor2009enthalpy}. Accordingly, the black hole mass $M$ will act as the enthalpy $H$ of the thermodynamic system rather than the internal energy. Then the first law has the following form
\begin{equation}
dH=T d\,S+V d\,P+\Phi d\,Q +\mathcal{A} d\,a,
\end{equation}
where $\mathcal{A}=\frac{\partial H}{\partial a}$ is the conjugate quantity to the QED parameter $a$. The entropy $S$, thermodynamic volume $V$, electric potential $\Phi$, and $\mathcal{A}$ can be calculated as
\begin{align}
S & = \int T^{-1} dH = \pi r_{+}^{2}, \\
V & = \left( \frac{\partial H}{\partial P}\right)_{S,Q}=\frac{4}{3}\pi r_+^3,  \\
\Phi & = \left( \frac{\partial H}{\partial Q} \right)_{S,P}=\frac{Q}{r_+}-\frac{a Q^3}{10 r_+^5}, \\
\mathcal{A}&= -\frac{Q^4}{40 r_{+}^{5}}. \label{Conja}
\end{align}
It is straightforward to confirm the following Smarr formula holds
\begin{equation}
 H=2(TS-VP+\mathcal{A}a)-\Phi Q.
\end{equation}
From Eq. (\ref{Hawkingtem}), the equation of state can be written as
\begin{equation}
P= \frac{T}{2r_+}-\frac{1}{8\pi r_+^2}+\frac{Q^2}{8\pi r_+^4}-\frac{a Q^4}{32\pi r_+^8}. \label{eos}
\end{equation}
In term of thermodynamic volume $V=\frac{4}{3}\pi r_+^{3}$, the equation of state reads
\begin{equation}
P=-\frac{\sqrt[3]{2} \pi ^{5/3} a Q^4}{9\times 3^{2/3} V^{8/3}}+\frac{\sqrt[3]{\frac{\pi }{6}} Q^2}{3V^{4/3}}+\frac{\sqrt[3]{\frac{\pi }{6}} T}{\sqrt[3]{V}}-\frac{1}{2\times 6^{2/3} \sqrt[3]{\pi } V^{2/3}}. \label{state}
\end{equation}
We plot the pressure $P$ as a function of the specific volume  $v=2\, r_+$ for fixed temperature in Fig. \ref{fig:Pvimage_1} with $a=-1.5$ ($a<0$) and $a=1.0$ ($0 \leq a \leq \frac{32}{7}Q^2 $) being two representative examples, respectively. In Fig. \ref{fig:VdWpv_1a}, we can see that there are two extremal points (marked with red dots) for $T<T_c$. And these points divide the isothermal curve into three branches. Two of them are the small black hole and large black hole locating at the left and right sides of the isothermal curve respectively, while the middle branch is for the intermediate black hole. According to the heat capacity, these black hole branches with negative slope are thermodynamic stable, while those with positive slope are unstable. Thus the small and large black holes are stable and the intermediate black hole is unstable. Making the use of the Maxwell equal area law, one can construct two equal areas along each isothermal curve to obtain the phase transition point. On the other hand, with the increase of the temperature, these two extremal points get closer and coincide with each other at $T=T_c$. When $T>T_c$, there is no extremal point anymore. Quite differently, for a positive QED parameter, i.e., $a=1.0\in(0 \leq a \leq \frac{32}{7}Q^2)$, there are three extremal points in Fig. \ref{fig:pv_1b} for $T<T_c$. When the temperature approaches its critical value, two of them coincide, while the other one continues to exist even when the temperature is above its critical value. Below the critical temperature, the nonmonotonic behavior of the isothermal curve also allows one to construct two equal areas indicating the existence of the black hole phase transition.

The critical point can be determined by the following conditions
\begin{equation}\label{critpoint}
\left( \partial_V P \right)_{Q,T,a} = \left( \partial_{V,V} P \right)_{Q,T,a}=0.
\end{equation}
These conditions gives a third degree equation for $x= \left(\frac{6}{\pi }\right)^{2/3} V^{2/3}$
\begin{equation}
x^3-24 Q^2 x^2 +448 a Q^4=0.\label{cubic}
\end{equation}
When $ 0\leq a \leq \frac{32}{7} Q^2 $, the equation (\ref{cubic}) has three real roots \cite{Magos:2020ykt}
\begin{equation}
x_k= 8Q^2 \left( 2\cos{\left[\frac{1}{3}\arccos{\left( 1-\frac{7a}{16Q^2}\right)-\frac{2\pi k}{3}}\right]}+1\right), \hspace{.5cm} k=0,1,2.
\label{xk}
\end{equation}
One notes that $x_2$ gives a negative volume, and it should be excluded. Therefore, only $k$=0 and 1 is allowed, and the critical temperature and pressure are given by
\begin{eqnarray}\label{critical_point}
T_{ck}&=&\frac{8\times 2^{2/3} \pi ^{4/3} a Q^4}{9 \sqrt[3]{3} V_c^{7/3}}-\frac{4 Q^2}{3 V_c}+\frac{1}{\sqrt[3]{6} \pi ^{2/3} V_c^{3/2}}, \nonumber \\
P_{ck}&=&\frac{7 \sqrt[3]{2} \pi ^{5/3} a Q^4}{9\times 3^{2/3} V_c^{8/3}}-\frac{\sqrt[3]{\frac{\pi }{6}} Q^2}{V_c^{4/3}}+\frac{1}{2\times 6^{2/3} \sqrt[3]{\pi } V_c^{2/3}},\hspace{.5cm}k=0,1.
\end{eqnarray}
The critical volume is defined as  $V_c=\frac{1}{6} \pi  x_{k}^{3/2}$ depending on the charge $Q$ and QED parameter $a$. It is worth noting that when the QED parameter $a > \frac{32 Q^2}{7}$, the equation (\ref{cubic}) has no real root, and thus no critical point exists. For $a<0$, there is a real root $x_0$ ($k$=0) giving one critical point.

In summary, there are two characteristic types of phase transition. The first type is for $a<0$, which admits one critical point, and the phase transition is the similar to the gas-liquid phase transition of VdW fluid. For the other type with QED parameter $0<a<\frac{32 Q^2}{7}$, two critical points can be found, and we shall see there exists the reentrant phase transition.

In the extended phase space, we can introduce the reduced quantities $ \tilde{P}$, $\tilde{T}$, $\tilde{V}$, $ \tilde{v} $
\begin{equation}
\tilde{P}=\frac{P}{P_c},\hspace{1cm} \tilde{T}= \frac{T}{T_c}, \hspace{1cm}
\tilde{V}=\frac{V}{V_c},\hspace{1cm}\tilde{v}=\frac{v}{v_c}. \label{reduced}
\end{equation}
Then the equation of state (\ref{state}) takes the following form
\begin{equation}\label{redeq}
	\tilde{P}=\frac{1}{\pi  P_c v_c^2}\left(-\frac{8 a Q^4}{\tilde{v}^8 v_c^6}+\frac{2 Q^2}{\tilde{v}^4 v_c}-\frac{1}{2\tilde{v}^2}\right)+\frac{\tilde{T}}{\tilde{v} \rho _c},
\end{equation}
where $ \rho_c=\frac{P_{c}v_{c}}{T_c} $. In the canonical ensemble, the Gibbs free energy $G=H-TS$, which reads
\begin{equation}
G(P,V,Q,a)=-\frac{7 \pi ^{5/3} a Q^4}{15\times 6^{2/3} V^{5/3}}-\frac{P V}{2}+\frac{3^{2/3} \sqrt[3]{\frac{\pi }{2}} Q^2}{2
	\sqrt[3]{V}}+\frac{\sqrt[3]{\frac{3}{\pi }} \sqrt[3]{V}}{4\times 2^{2/3}}.
\label{Gibbs}
\end{equation}
In general, the first-order phase transition can be determined by the swallow tail behavior of the Gibbs free energy. In the following sections, we shall study the phase transition via the behavior of the Gibbs free energy.


\subsection{Van der Waals type phase transition with $a<0$}

\begin{figure}[h]
\centering
\subfigure[]{\includegraphics[width=7cm]{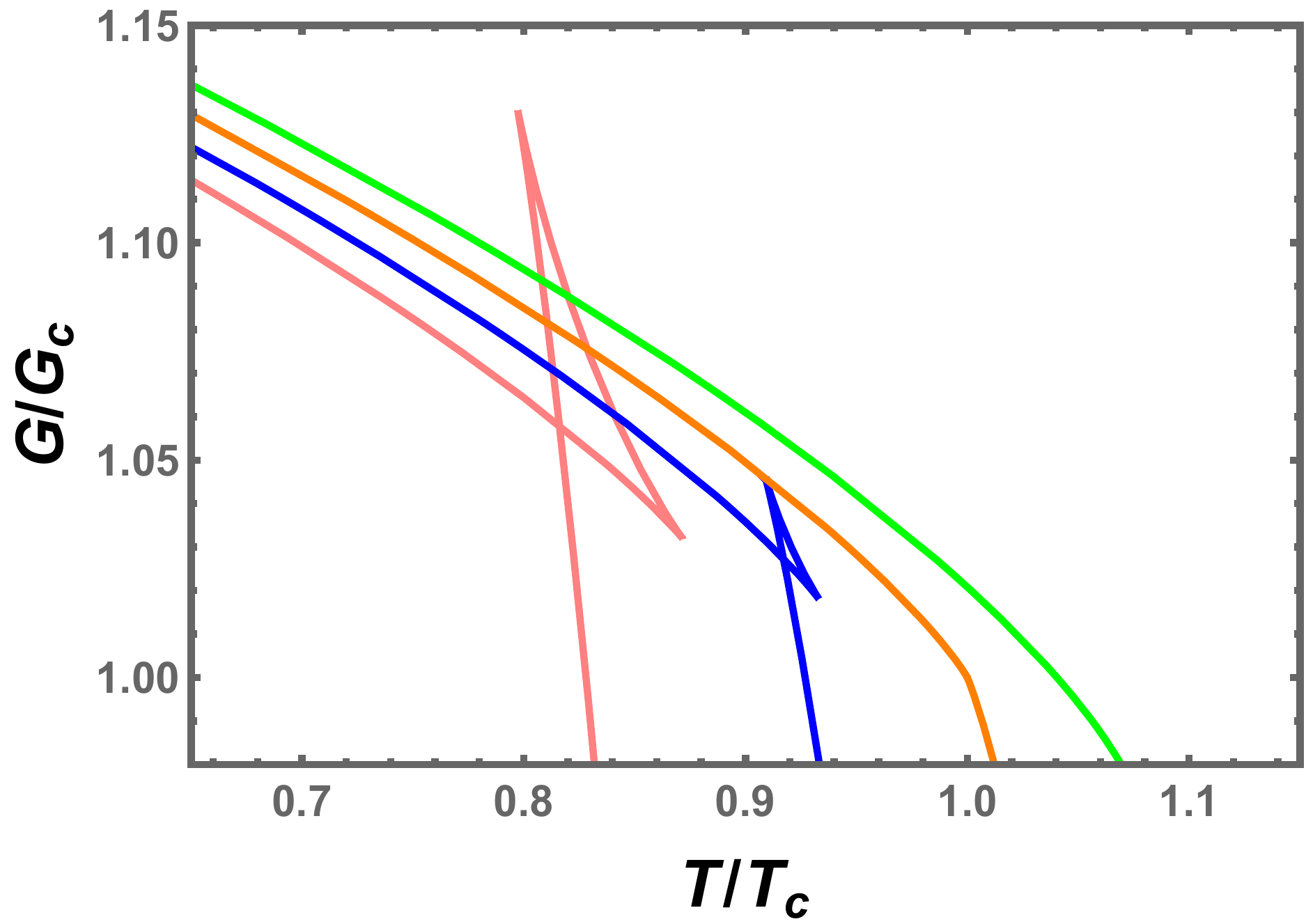}\label{fig:vdwgt_2a}   }
\subfigure[]{\includegraphics[width=7cm]{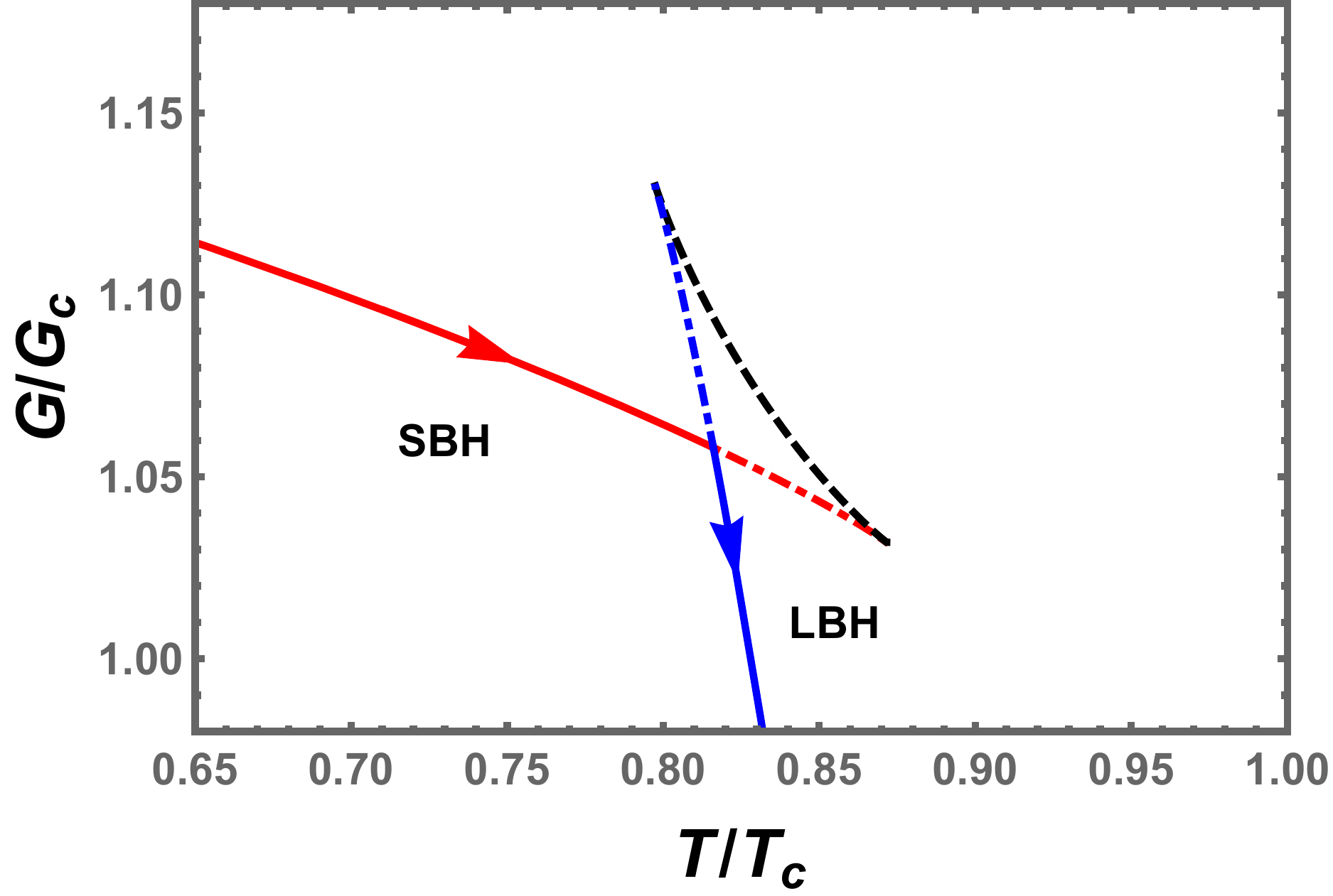}\label{fig:vdwgt_2b}}
\caption{(a) The Gibbs free energy for $\tilde{P}=0.6 $ (pink solid line), $ \tilde{P}=0.8  $ (blue solid line), $ \tilde{P}=1.0 $ (orange solid line) and $ \tilde{P}=1.2 $ (green solid line). (b) The small/large black hole phase transition. The SBH and LBH hole are marked with a red solid line and a blue solid line respectively and are stable, but the black hole marked with a black dashed line is unstable. Although the red and blue dotdashed lines are stable, they are not the minimum value of Gibbs free energy. Along the direction of the arrow, the volume of the black hole increases. We have set $ Q=1.0 $ and $ a=-1.5 $. }
\label{fig:vdwGT_2}
\end{figure}

For this case, we take $Q=1.0$ and $a=-1.5$ for example. The Gibbs free energy $G$ is plotted as the temperature $T$ in Fig. \ref{fig:vdwGT_2} for different values of the pressure. When $P<P_c$, we observe the characteristic swallow tail behavior indicating the coexistence of the black hole phase transition. As the pressure increases, the shape of the swallowtail becomes smaller, and it shrinks to a point when $P=P_c$. Further increasing the temperature, the behavior completely disappears. Then the Gibbs free energy turns to a smooth function of the temperature.

\begin{figure}[h]
\centering
\subfigure[]{ \includegraphics[width=7cm]{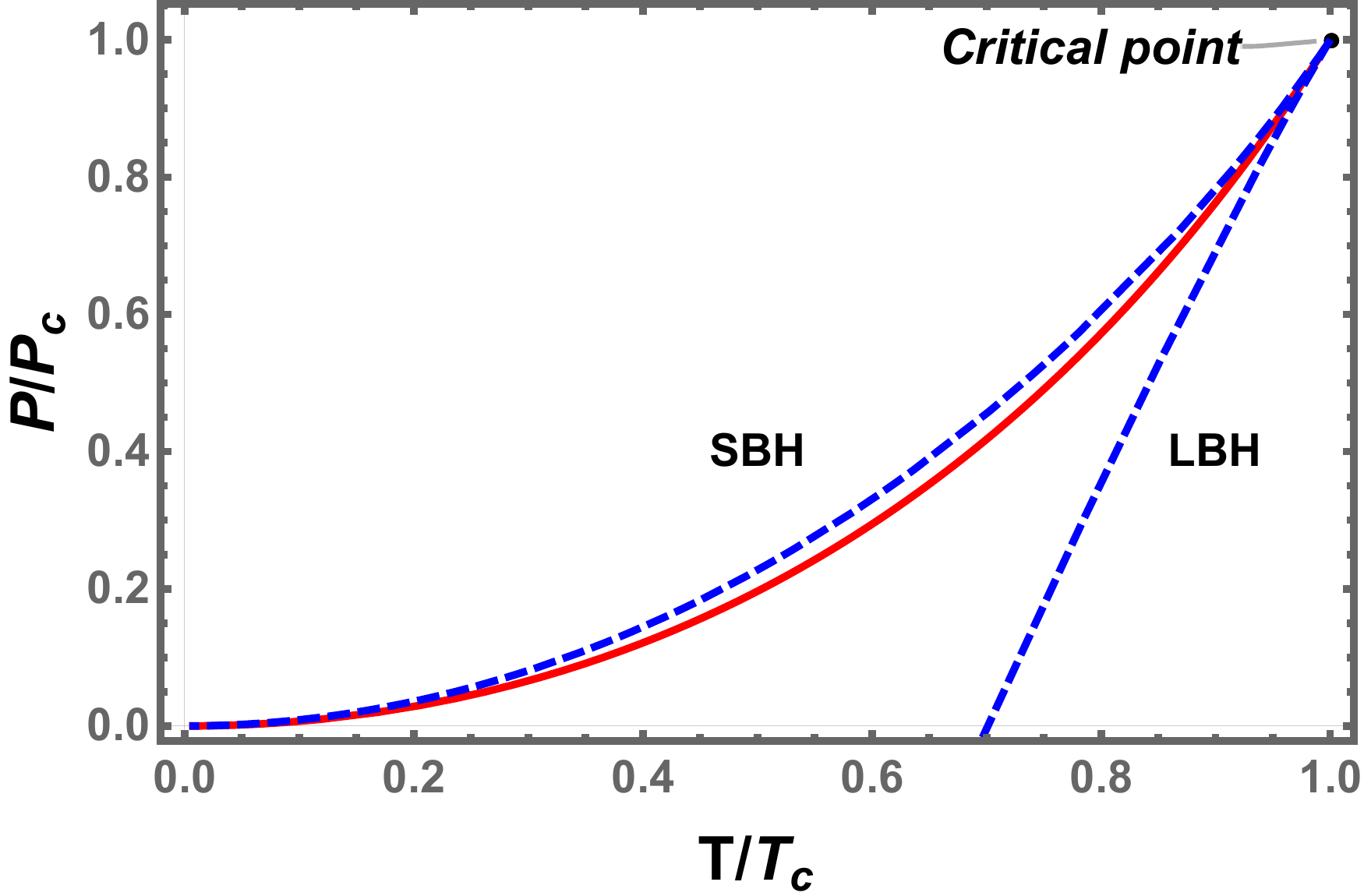}\label{fig:VdWPTphase}  }
\subfigure[]{ \includegraphics[width=7cm]{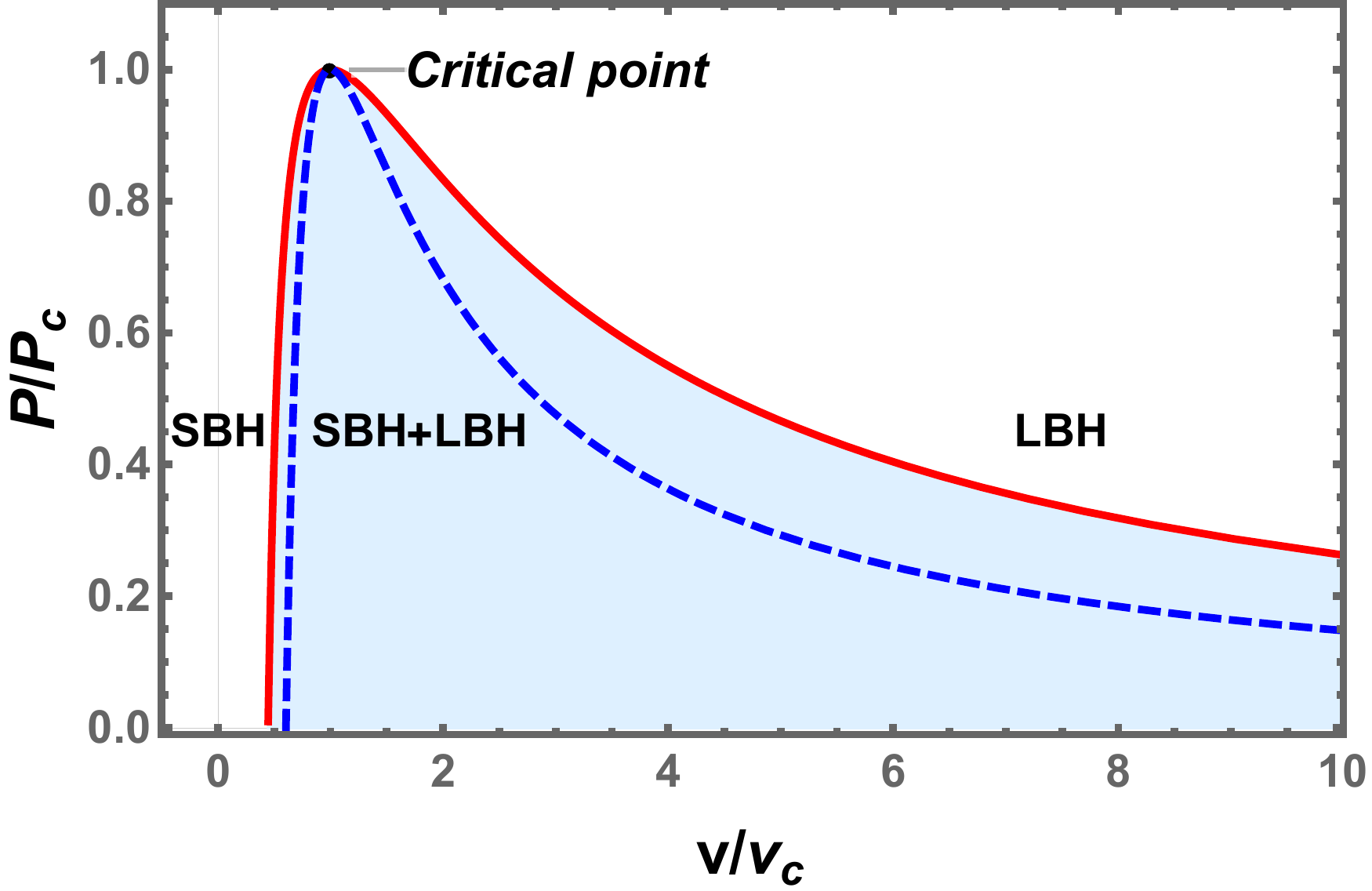}\label{fig:VdWPVphase} }
\caption{Phase diagram for VdW phase transition type case. (a) First-order coexistence curve (solid red) and spinodal curve (blue dashed line) are shown in the $P$-$T$ plane. The coexistence curve separates SBH and LBH phases and ends at a critical point. (b) The phase structure in the $ P $-$ v $ plane. The shadow region under the coexistence curve is the coexistence phase of the small and large black holes. The region between the coexistence curve and spinodal curves is metastable. Here we have taken $ Q=1.0 $ and $ a=-1.5 $.}
\label{fig:VdWphasestruct_3}
\end{figure}

In order to clearly show the phase transition, we describe the Gibbs free energy in Fig. \ref{fig:vdwgt_2b} with $P<P_c$. After a simple calculation, we find that the small and large black hole branches in blue and red solid curves have positive heat capacity, and thus they are thermodynamic stable. Whereas these branches in dashed curve are unstable or metastable. Considering that a system always prefers a state of low Gibbs free energy, the system will undergo a first-order phase transition from a small black hole (SBH) phase to a large black hole (LBH) phase with the increase of the temperature. The phase transition point exactly locates at the intersection of the swallow tail behavior. Thus it is easy to see that this phase transition is similar to the liquid-gas phase transition of a VdW fluid.

The phase structures of the charged EH-AdS black hole are given in Fig. \ref{fig:VdWphasestruct_3}, from which we find that it shares a similar phase diagram with the VdW fluid. In the $P$-$T$ diagram, the first-order coexistence curve of the small and large black holes starts at $ \tilde{T}=0 $ and ends at the critical point, which divides the plane into two regions corresponding to the small and large black hole phases, respectively. Considering that the equation of state is not applicable in the shadow region, we use the spinodal curve marked with the blue dashed curve to distinguish the metastable phase from the coexistence phase of the black hole. The spinodal curve is determined by
\begin{equation}
(\partial_V P)_T=0 \hspace{0.5cm}\text{or}\hspace{0.5cm} (\partial_V T)_P=0.
\end{equation}
Obviously, the spinodal curve meets the coexistence curve at the critical point.

\subsection{Reentrant phase transition case with $0\leq a\leq \frac{32 Q^2}{7}$}

For this case, two critical points can be found. The corresponding phase transition is the reentrant phase transition including a zeroth-order and a first-order phase transitions at a certain region of the temperature or pressure. Here we shall study the reentrant phase transition for the charged EH-AdS black hole.

\begin{figure}[h]
\centering
\includegraphics[width=8cm]{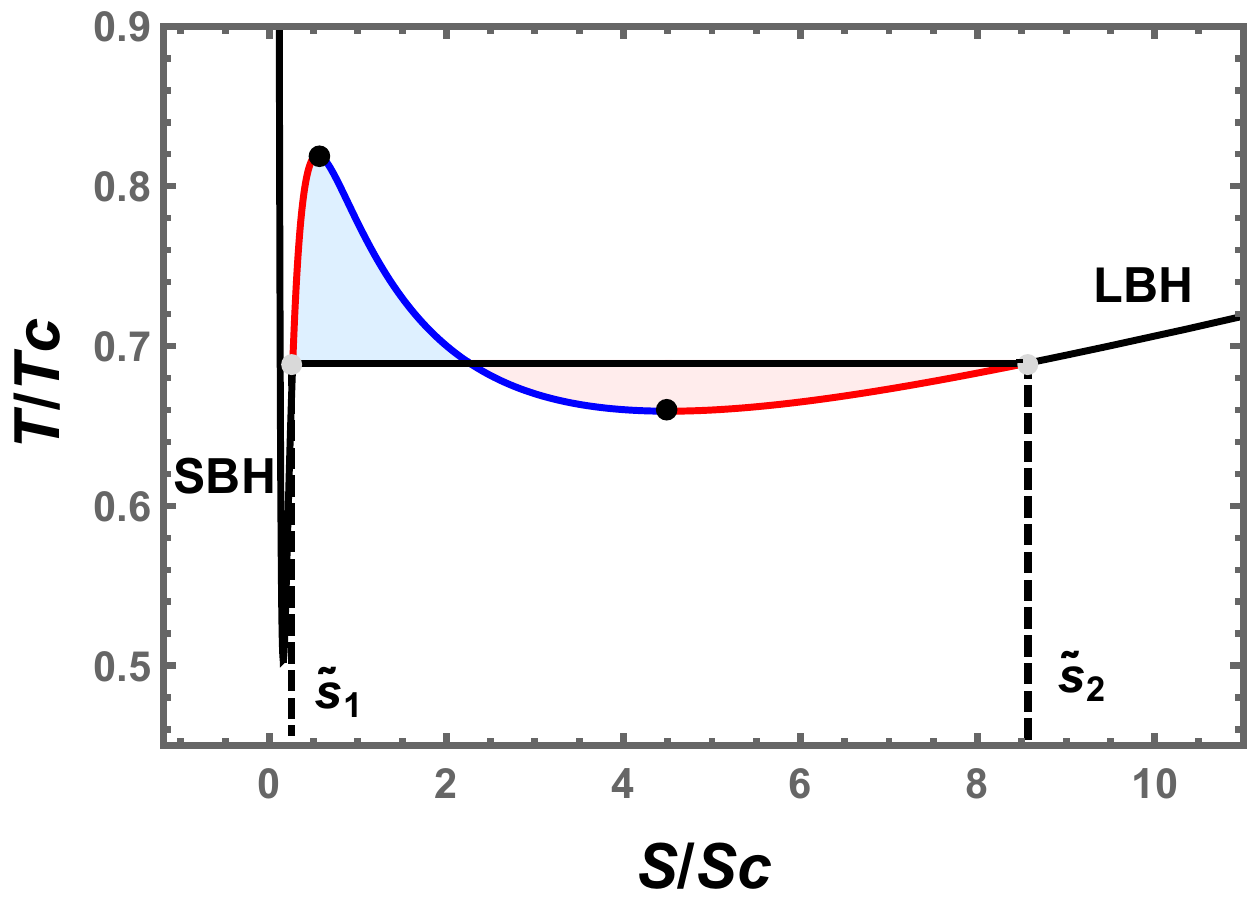}
\caption{Isobaric curve with $\tilde{P}=0.4 $ and the equal area law in the $\tilde{T}$-$\tilde{S}$ plane, where the two shadow areas above and below the horizontal line are equal. The black curves represent the small and large black hole branches respectively. These two red curves stand for the superheated SBH branch and the supercooled LBH branch. The blue curve decreases an unstable branch. We have set $ Q=1.0 $ and $ a=1.0 $.}
\label{fig:eqarea_4}
\end{figure}

As expected, the first-order phase transition can be determined by constructing two equal areas according to the Maxwell's equal area law. As we shown in Fig. \ref{fig:VdWpv_1a}, the isotherm curves exhibit a nonmonotonic behavior in $P$-$v$ plane. So we can construct two equal areas between two stable black hole branches. However, as pointed out in Ref. \cite{wei2015clapeyron}, the equal areas should be constructed in $P$-$V$ plane rather than $P$-$v$ plane. Alternatively, the Maxwell's equal area law also holds in the $\tilde{T}$-$\tilde{S}$ plane for giving pressure both in ordinary or reduced parameter space
\begin{align}
\tilde{T}_0 (\tilde{S}_2&-\tilde{S}_1)
=\int_{\tilde{S}_1}^{\tilde{S}_2} \widetilde{T} (\tilde{S}){\rm d}\tilde{S}, \, \qquad
\tilde{T}_0=\tilde{T} (\tilde{S}_1)=\tilde{T} (\tilde{S}_2),
\end{align}
where $\tilde{T}_0$ denotes the reduced temperature of phase transition and $\tilde{S}_1$ and $\tilde{S}_2$ are the reduced entropy of the corresponding coexistence small and large black holes. Taking $\tilde{P}$=0.4 as an example, the Maxwell's equal area law is fulfilled in the $\tilde{T}$-$\tilde{S}$ plane in Fig. \ref{fig:eqarea_4}. The black curves denote the small black hole and large black hole branches, respectively. The red curves are for the two metastable branches, the superheated small black hole branch and the supercooled large black hole branch. The blue curve with a negative slope is an unstable branch, and it is substituted by a horizontal line according to the equal area law. The area under the isobaric curve from $ \tilde{S}_{1} $ to $ \tilde{S}_{2} $ is required to equal the area enclosed by the rectangle under the isothermal horizontal line, so the areas of the two shadow regions are equal. These two black dots are the spinodal points, which separate the metastable branches from the unstable branch. Then the pressure and temperature corresponding to the isothermal horizontal line are just that of the phase transition point.

\begin{figure}
\begin{center}
	\subfigure[]{\includegraphics[width=7cm]{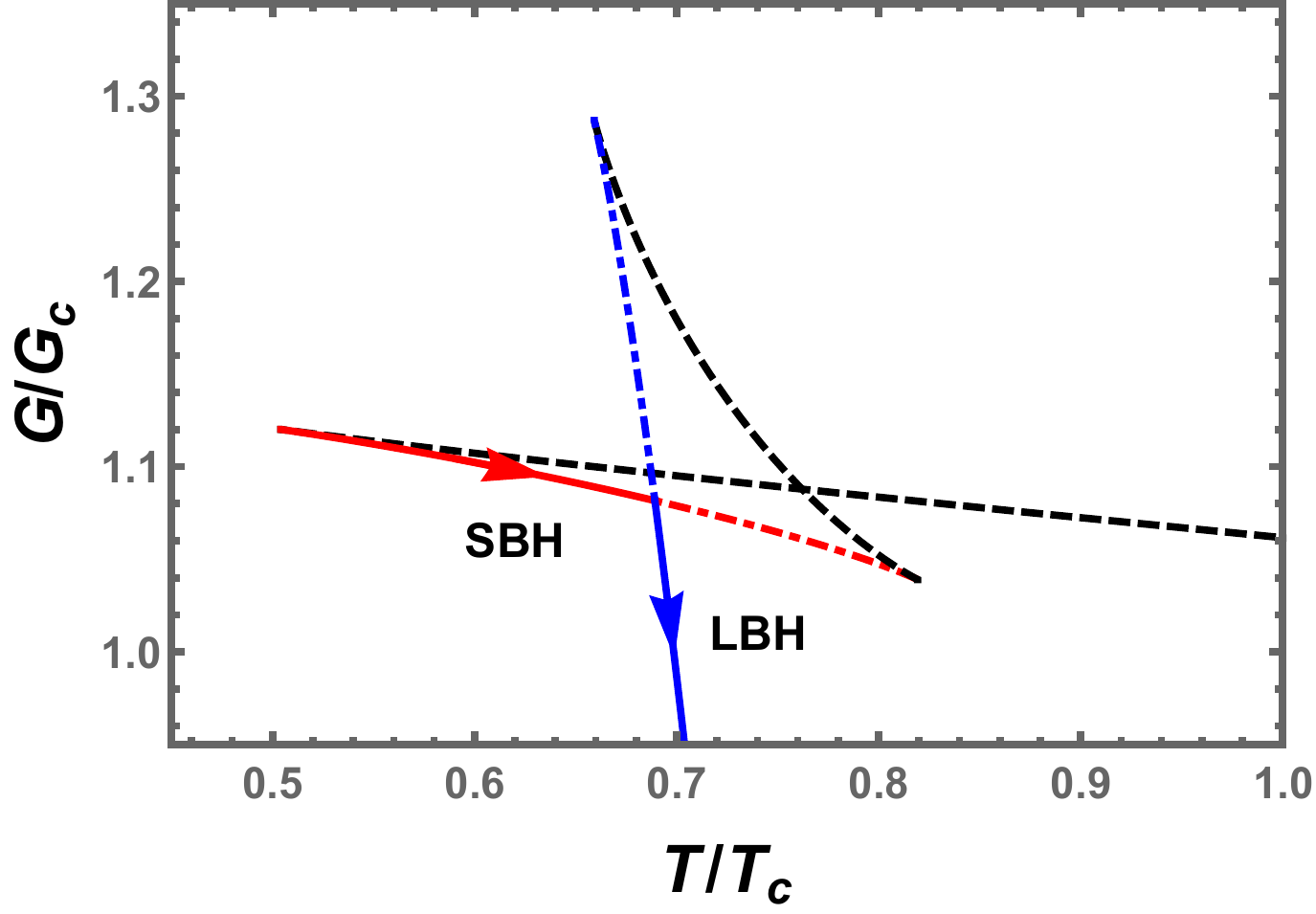}\label{fig:slpt_5a}}
	 \subfigure[]{\includegraphics[width=7cm]{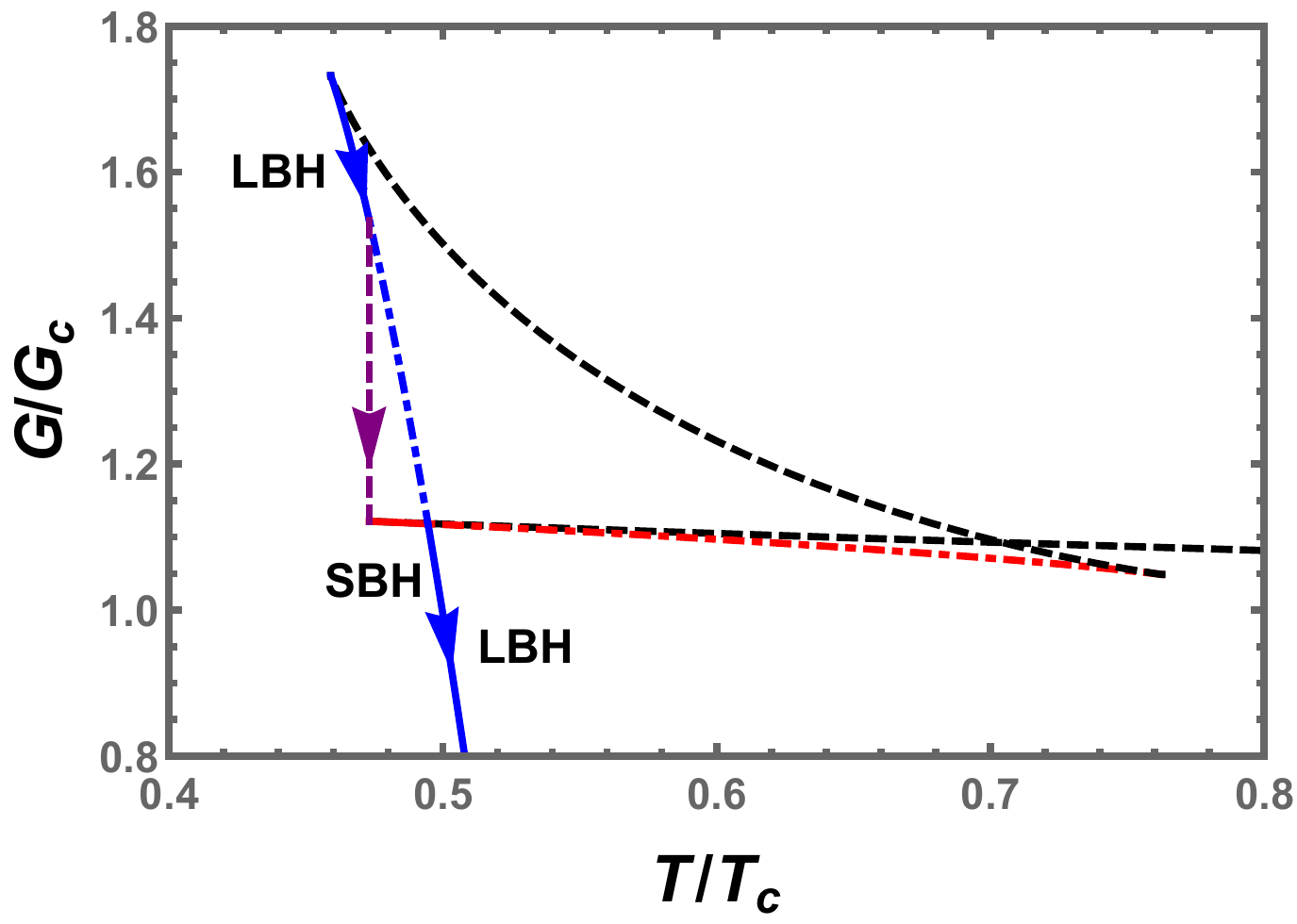}\label{fig:reentrant_5b}}
\end{center}
\caption{(a) The small/large black hole phase transition with $\tilde{P}=0.40$. (b) Reentrant phase transition (large/small/large black hole phase transition) with $\tilde{P}=0.19$. The SBH (red solid line) and LBH (blue solid line) are stable, while the black dashed curves are unstable. It is worth noting that the red and blue dot dashed curves correspond positive heat capacity. However they are not the minimum value of Gibbs free energy, and thus are metastable. We have set $ Q=1.0 $ and $ a=1.0 $.}
\label{fig:PTana_5}
\end{figure}

We depict the Gibbs free energy for these two types of phase transitions in Fig. \ref{fig:PTana_5}. In Fig. \ref{fig:slpt_5a}, the small/large black hole phase transition is analyzed for the reduced pressure $ \tilde{P}=0.4$. As the temperature increases, the system jumps directly from the SBH phase (red solid line) to the LBH phase (blue solid line), and the volume of the black hole directly undergoes a sudden change when the temperature is equal to the phase transition temperature. For the black hole branches described by the black dashed lines, their heat capacity is negative indicating thermodynamically instability. Note that although these black hole branches described by the dotdashed lines in the Fig. \ref{fig:PTana_5} have a positive heat capacity, they do not have the lowest free energy for a fixed temperature, and thus they are metastable black hole branches. In addition to the small/large black hole phase transition, another new zeroth-order phase transition emerges in Fig. \ref{fig:reentrant_5b}. When the zeroth-order phase transition occurs, the black hole system jumps from the LBH phase to the SBH phase. Moreover, not only the volume of a black hole will undergo sudden changes, but the Gibbs free energy will also change drastically. In short, the black hole system firstly undergoes a zeroth-order phase transition from LBH phase to SBH phase and then returns to LBH phase through the small/large black hole phase transition with the increase of the temperature. This phase transition of such pattern is called the reentrant (large/small/large BH) phase transition.

\begin{figure}[h]
\centering
\subfigure[]{ \includegraphics[width=7cm]{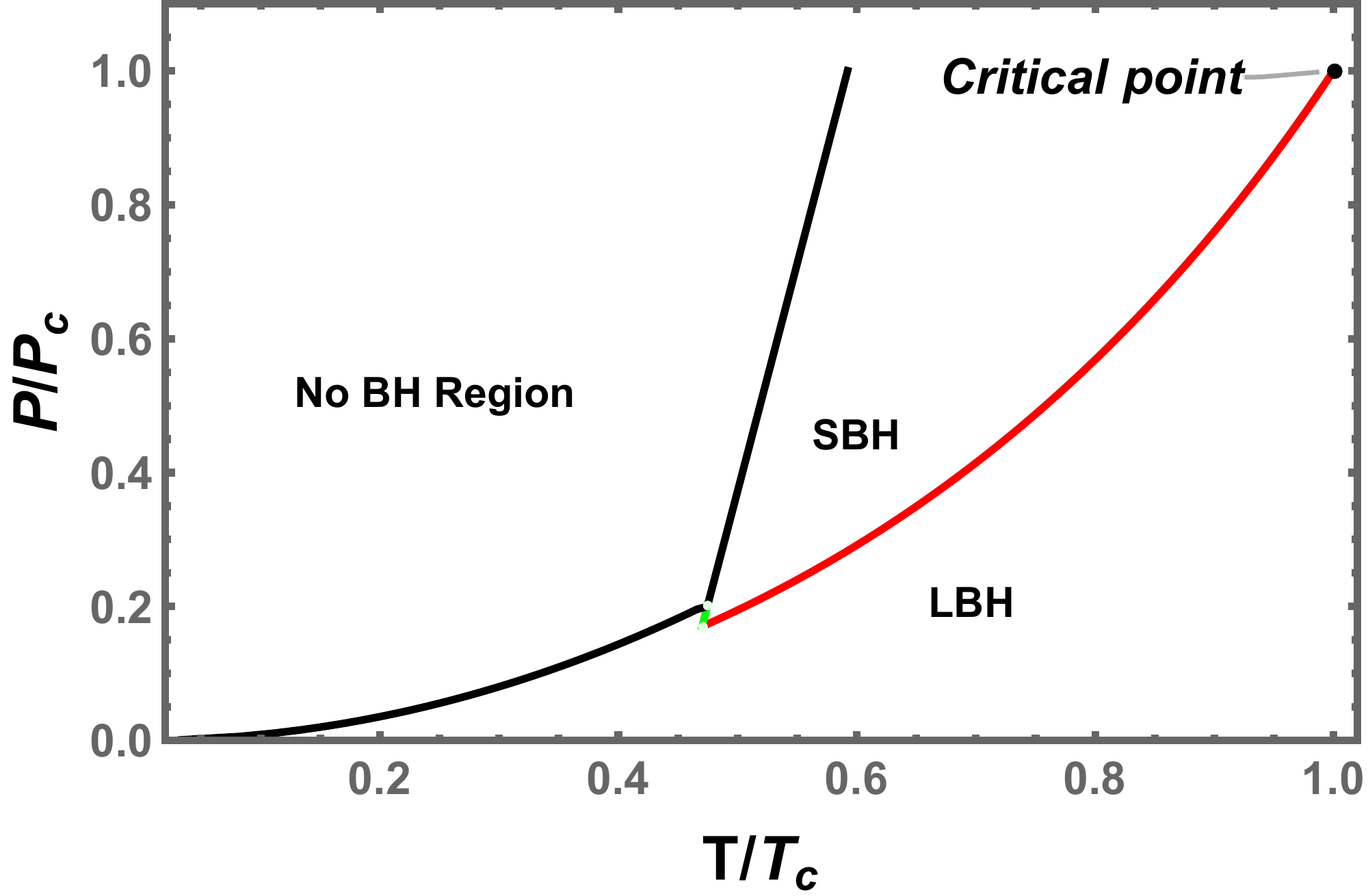}\label{fig:coexistenPT_6a}  }
\subfigure[]{ \includegraphics[width=7cm]{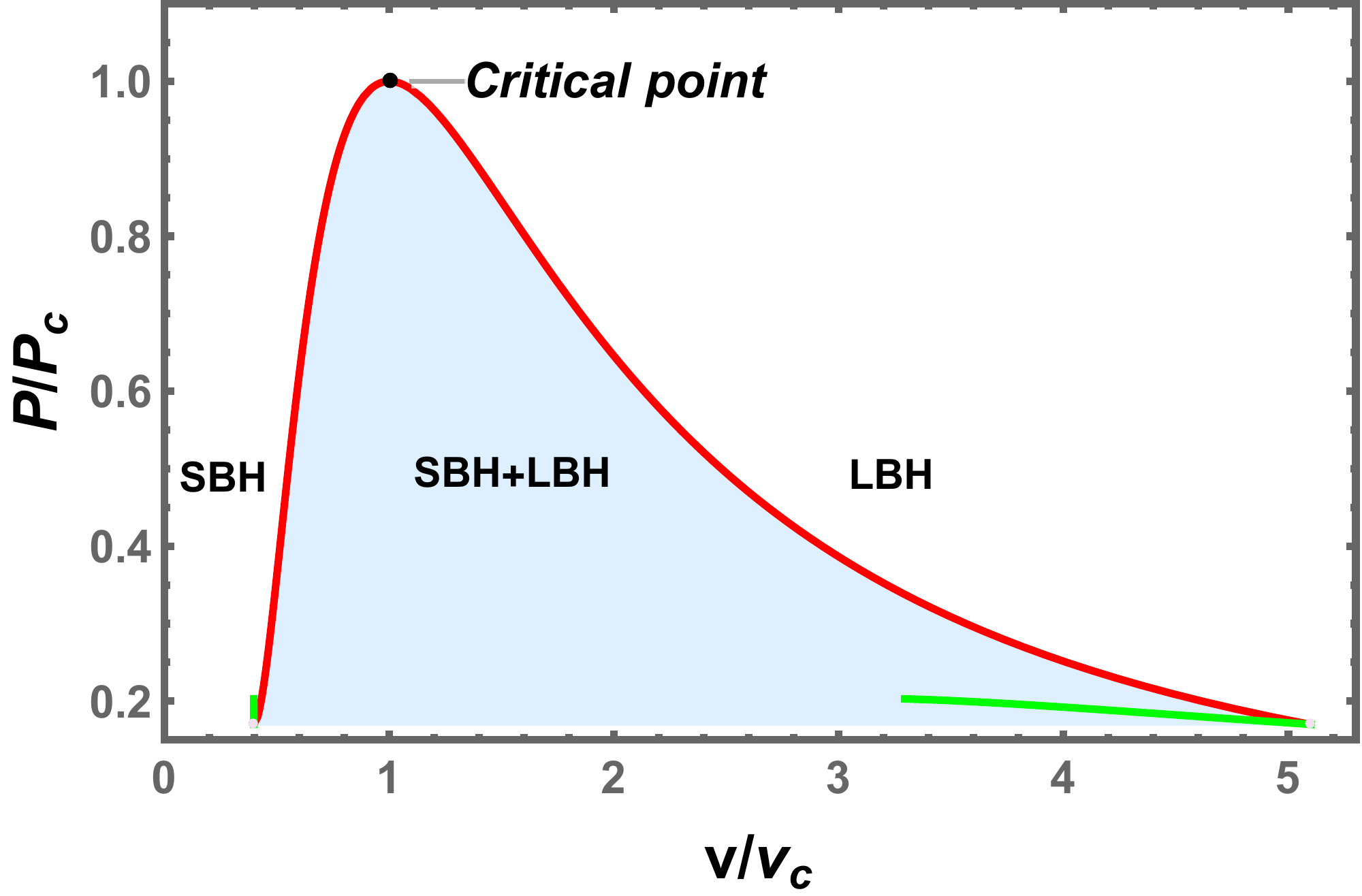}\label{fig:coexistenPV_6b} }
\caption{Phase diagram for the charged EH-AdS black hole. (a) The first-order coexistence curve (solid red) and zeroth phase transition line (solid green) are also shown in the $\tilde{P}$-$\tilde{T}$ plane. Black line separates no black hole region from black hole region. The first-order coexistence curve separates SBH and LBH phases and ends at a critical point. (b) The phase structure in the $\tilde{P}$-$\tilde{v}$ plane. The shadow area under the coexistence curve is the coexistence phase of the small and large black holes.}
\label{fig:phasediagram_6}
\end{figure}

The phase diagrams of reentrant phase transition are shown in Fig. \ref{fig:phasediagram_6}. Similar to the case of the VdW phase transition type, the coexistence curve (red solid line) divides the parameter space into two regions in Fig. \ref{fig:coexistenPT_6a}. Above and below the coexistence curve are the SBH phase and the LBH phase, respectively. The zeroth-order phase transition line marked with green solid line connects the first-order coexistence curve and the minimum temperature curve. The phase diagram is also exhibited in $\tilde{P}$-$\tilde{v}$ plane in Fig. \ref{fig:coexistenPV_6b}. The critical point divides the coexistence curve into left and right parts corresponding to the coexistence small black hole and the large black hole, respectively. The small and large black hole regions locate at the left and right.

\begin{figure}[]
\centering
\subfigure[]{ \includegraphics[width=7cm]{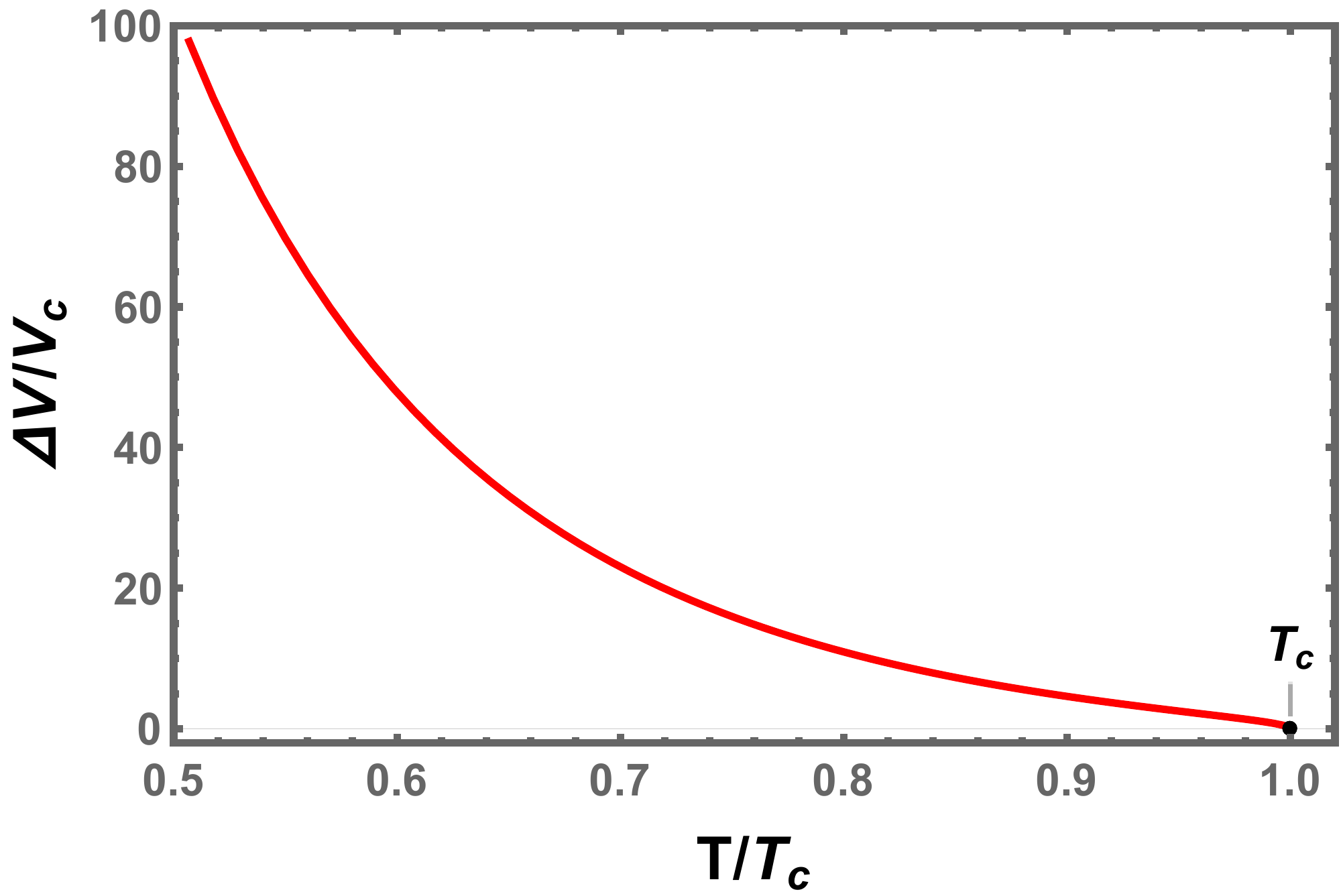}\label{fig:deltaVT_7a}  }
\subfigure[]{ \includegraphics[width=7cm]{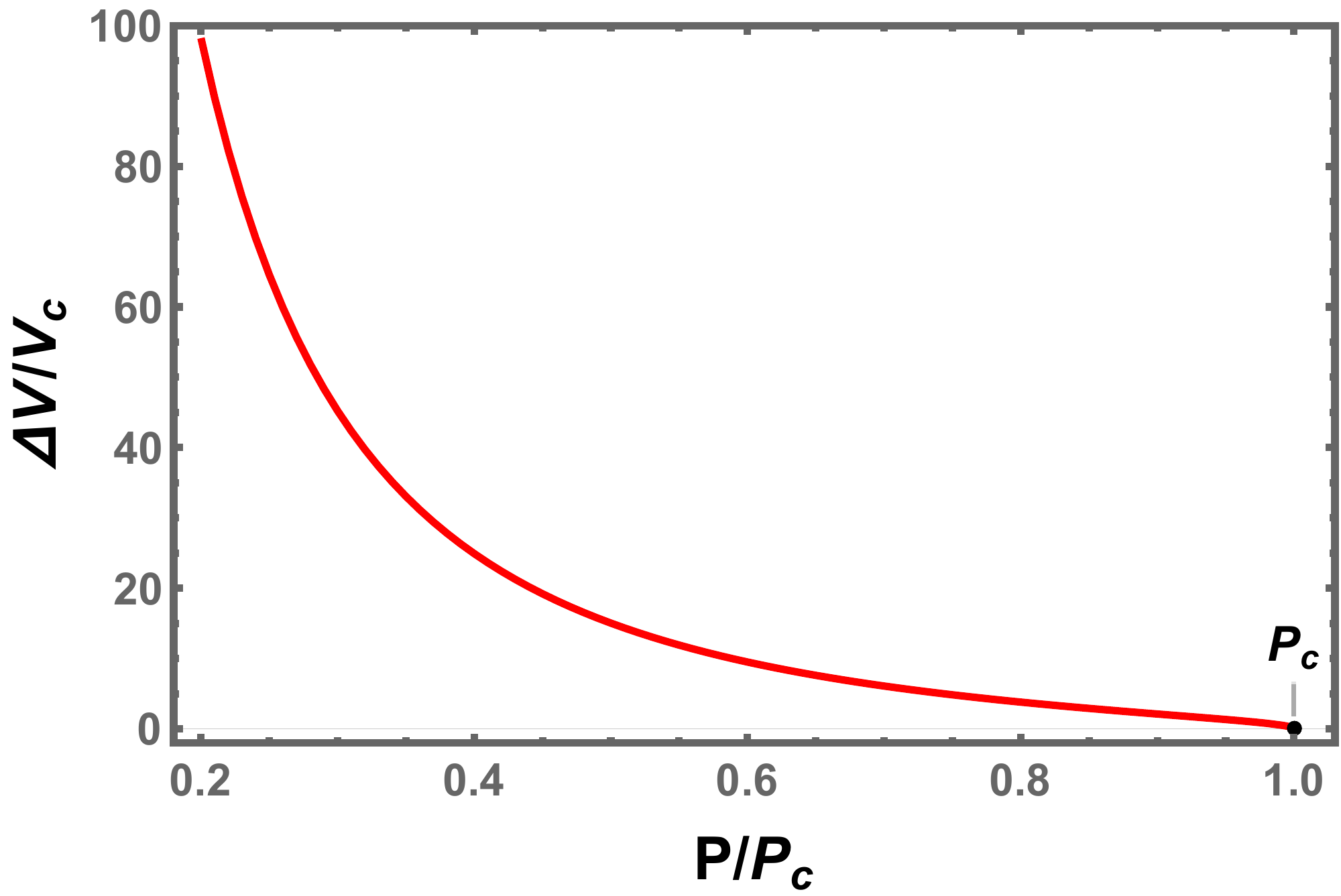}\label{fig:deltaVP_7b} }
\caption{The behavior of the change of the thermodynamic volume at the black hole phase transition. (a) $\Delta V  $ vs. $ \tilde{T} $. (b) $ \Delta V $ vs. $ \tilde{P} $. The critical point is at $ T_c $ and $ P_c $.}
\label{fig:del_7}
\end{figure}

In Fig. \ref{fig:del_7}, we plot the difference of the volume $\Delta V=V_{l}-V_{s}$ among the first-order phase transition as a function of the temperature and pressure, respectively. It is obvious that below the critical point, $\Delta V$ has a finite value indicating a sudden change during the black hole phase transition. However, when the critical point is reached, $\Delta V$ vanishes, which means that one could not distinguish the small and large black holes anymore. Considering the behavior of $\Delta V$, it can be regarded as an order parameter to characterize the phase transition.

Near the critical point, the critical exponent of $\Delta V$ can be calculated, see Ref. \cite{Magos:2020ykt}. However, in the calculation, the specific volume $v$ is used in constructing the equal area law, which as we pointed out is inappropriate. Here we would like to make use of the thermodynamic volume $V$ instead, and to see whether the result keeps unchanged. Using the relationship between thermodynamic volume and specific volume $V=\frac{\pi v^3}{6}$, we get $\tilde{V}=\tilde{v}^{3}$. Then the formula (\ref{redeq}) can be expressed as
\begin{align}\label{tildep}
	\tilde{P}&=\frac{1}{\pi  P_c v_c^2}\left(\frac{2 Q^2}{\tilde{V}^{4/3} v_c}-\frac{8 a Q^4}{\tilde{V}^{8/3} v_c^6}-\frac{1}{2\tilde{V}^{2/3}}\right)+\frac{\tilde{T}}{\tilde{V}^{1/3} \rho _c},
\end{align}
where $ \rho_c=\frac{P_{c}v_{c}}{T_c} $. Let the first term in the previous equation be the function $ h(\tilde{V}) $, so (\ref{tildep}) can be written as
\begin{equation}
\tilde{P}(\tilde{V},\tilde{T})=h(\tilde{V})+\frac{\tilde{T}}{\tilde{V}^{1/3} \rho _c}.
\end{equation}
The Taylor expansion in the vicinity of the critical point at $ \tilde{V}=1 $ and $ \tilde{T}=1 $ is written as
\begin{align}\label{taylorexp}
	\tilde{P}(\tilde{T},\tilde{V}) \simeq  \nonumber&\frac{1}{\rho _c}+h(1)+(\tilde{V}-1) \left(h'(1)-\frac{1}{3 \rho _c}\right)+(\tilde{V}-1)^2 \left(\frac{2}{9 \rho
	_c}+\frac{h''(1)}{2}\right)\\
&+(\tilde{V}-1)^3 \left(\frac{1}{6} h^{(3)}(1)-\frac{14}{81
	\rho _c}\right)+(\tilde{T} -1) \left(\frac{1}{\rho _c}-\frac{\tilde{V}-1}{3 \rho _c}+\frac{2
	(\tilde{V}-1)^2}{9 \rho _c}\right).
\end{align}
From the definition of the critical point, the equation (\ref{taylorexp}) must meet the following conditions
\begin{align}
\nonumber \tilde{P}(\tilde{V},\tilde{T})\bigg|_{\tilde{V}=1,\tilde{T}=1}&   =\frac{1}{\rho_c}+h(1)=1,\\
\nonumber \partial_{\tilde{V}} \tilde{P}(\tilde{V},\tilde{T}) \bigg|_{\tilde{V}=1,\tilde{T}=1}&= h'(1)-\frac{1}{3\rho _c}=0,\\
\partial_{\tilde{V},\tilde{V}} \tilde{P}(\tilde{V},\tilde{T}) \bigg|_{\tilde{V}=1,\tilde{T}=1}& =\frac{4}{9\rho _c}+h''(1)=0.
\end{align}
Therefore, Eq. (\ref{taylorexp}) can be simplified to
\begin{equation}
	\tilde{P}(\tilde{T},\tilde{V}) =1 + \left(\tilde{V}-1\right)^3 \left(\frac{1}{6} h^{(3)}(1)-\frac{14}{81\rho _c}\right)+\left(\tilde{T}-1\right) \left(\frac{1}{\rho _c}-\frac{\tilde{V}-1}{3 \rho _c}+\frac{2(\tilde{V}-1)^2}{9 \rho _c}\right).  \\
\end{equation}
Taking $ \omega=\tilde{V}-1 $ and $ t=\tilde{T}-1$, the previous expression become
\begin{equation}
	\tilde{P}(t,\omega)=1+t \left(\frac{1}{\rho _c}-\frac{\omega }{3\rho _c}\right)-C \omega ^3+  \mathcal{O}(t\omega^2,\omega^4),
\end{equation}
where we have defined $ C=\frac{14}{81\,\rho _c}-\frac{1}{6} h^{(3)}(1) $. From Maxwell's equal area law we obtained
\begin{equation}\label{eq29}
\int_{P_s}^{P_l} V\,dP=\int_{\omega_s}^{\omega_l} P_c V_c\left(\omega+1 \right) \left(\frac{t}{3\rho _c}+3 C \omega ^2\right)\,d\omega=0,
\end{equation}
where $\omega_l$ and $\omega_s$ are the volumes of the coexistence small and large black hole respectively. For an isothermal process, the pressure $ P_s $ is equal to $ P_l $, i.e. $\tilde{P}_s(t,\omega)=\tilde{P}_l(t,\omega)$, which reads
\begin{equation}\label{eq30}
\frac{ t\,\omega _l}{3\rho _c}+C \omega _l^3= \frac{t\,\omega _s}{3\rho _c}+C \omega _s^3.
\end{equation}
Combining (\ref{eq29}) and (\ref{eq30}), we have
\begin{equation}
\omega_l=-\omega_s=\sqrt{\frac{ - 3}{5 C\, \rho_c}} \sqrt{t}.
\end{equation}
The critical exponent $ \beta $ is related to the order parameter $ \eta = |t|^{\beta} $, namely with the change of volume at the phase transition for a given isotherm process,
\begin{equation}
\eta = V_l-V_s=V_c (\omega_l-\omega_s) =2\,V_c \sqrt{\frac{ - 3}{5 C\, \rho_c}} \sqrt{t}\propto |t|^{1/2}.
\end{equation}
Therefore the critical exponent $\beta$ is
\begin{equation}
\beta=\frac{1}{2}.
\end{equation}
As a result, we can see that the choose of the specific volume or thermodynamic volume does not change the value of the critical exponent $\beta$. However when applying the Maxwell equal area law, we should choose the thermodynamic volume rather than the specific volume.

\section{Ruppeiner geometry and microstructure}\label{c}

In this section, we would like to construct the Ruppeiner geometry for the charged EH-AdS black hole. Via the curvature scalar, the microstructure will be tested.

\subsection{Ruppeiner geometry}

Here, we firstly give a brief introduction for the Ruppeiner geometry and then calculate the curvature scalar for the charged EH-AdS black hole.

Let us consider a thermodynamically isolated system in equilibrium with the total entropy $S$. The system is divided into two subsystems, the small system we consider and its large environment. Their entropies are denoted by $S_S$ and $S_E$ with $S_{S}\ll S_{E}\sim S$. We suppose that the total entropy of the system is described by two independent thermodynamic variables $x_{0}$ and $x_{1}$. The total entropy of the system can be written as
\begin{equation}
S(x_0,x_1)=S_S(x_0,x_1)+S_E(x_0,x_1). \nonumber
\end{equation}
For an equilibrium system, the entropy reaches its maximum locally. We perform Taylor expansion in the neighborhood of this local maximum $(x^{\mu}=x^{\mu}_{0})$
\begin{equation}
	S=S_0+\frac{\partial S_S}{\partial x^\mu}\bigg|_{x^\mu_0}\Delta x^\mu_S
+\frac{\partial S_E}{\partial x^\mu}\bigg|_{x^\mu_0}\Delta x^\mu_E + \frac{1}{2}\frac{\partial^2 S_S}{\partial x^\mu \partial x^\nu}\bigg|_{x^\mu_0}\Delta x^\mu_S \Delta x^\nu_S
+\frac{1}{2}\frac{\partial^2 S_E}{\partial x^\mu \partial x^\nu}\bigg|_{x^\mu_0}\Delta x^\mu_E \Delta x^\nu_E
+\cdots,
\end{equation}
where a zeroth-order term $S_{0}$ is the local maximum of entropy at $x^{\mu}_{0}$. The entropy of an isolated system in equilibrium is conserved under virtual change. This shows that the first derivative of the entropy vanishes, and thus we get
\begin{align}
\Delta S=S-S_0&=\frac{1}{2}\frac{\partial^2 S_S}{\partial x^\mu \partial x^\nu}\bigg|_{x^\mu_0}\Delta x^\mu_S \Delta x^\nu_S
+\frac{1}{2}\frac{\partial^2 S_E}{\partial x^\mu \partial x^\nu}\bigg|_{x^\mu_0}\Delta x^\mu_E \Delta x^\nu_E
+\cdots    \nonumber  \\
&\approx \frac{1}{2}\frac{\partial^2 S_S}{\partial x^\mu \partial x^\nu}\bigg|_{x^\mu_0}\Delta x^\mu_S \Delta x^\nu_S.\label{sbbb}
\end{align}
It is worth noting that $S_{E}$ is a thermodynamic extensive quantity, and it is the same order of magnitude as the entropy of the entire system. Therefore, its derivative with respect to the intensive quantity $ x^{\mu} $ is much smaller than the derivative of $ S_{S} $, which can be ignored. Therefore, the
probability of finding the system in the internals ($x_0$, $x_0+dx_0$) and ($x_1$, $x_1+dx_1$) will be
\begin{align}
 P(x_0, x_1)\propto e^{\frac{\Delta S}{k_B}}=e^{-\frac{1}{2}\Delta l^2},
\end{align}
where $k_B$ is Boltzmann's constant. With (\ref{sbbb}), the line element of Ruppeiner geometry measuring the distance between two neighboring fluctuation states can be written as
\begin{align}\label{Rline}
\Delta l^2=&\frac{1}{k_B}g^R_{\mu \nu} \Delta x^\mu \Delta x^\nu,\\
g^R_{\mu \nu}&=-\frac{\partial^2 S_B}{\partial x^\mu \partial x^\nu}.
\end{align}
Because $ \Delta l^2 $ can measure the distance between two neighboring fluctuation states, the thermodynamic metric $ g_{\mu\nu}^{R} $ potentially contains some information about the microstructure of the system.

When we choose the thermodynamic coordinates $x^{\mu}$ to be the temperature $T$ and the volume $V$, the Helmholtz free energy is the thermodynamic potential. The corresponding line element can be written as follows \cite{Wei:2019uqg}
\begin{equation}
\Delta l^2=\frac{C_V}{T^2} \Delta T^2-\frac{(\partial_V P)_T}{T} \Delta V^2,    \label{IM}
\end{equation}
where $ C_{V}=T(\partial_TS)_{V} $ is the heat capacity at constant volume. Using the convention in the literature \cite{ruppeiner1995riemannian}, we can directly calculate the corresponding scalar curvature of the line element
\begin{align}
R&=\frac{1}{2C_V^2(\partial_V P)^2}\bigg \{ T(\partial_V P) \Big[ (\partial_T C_V)(\partial_V P-T\partial_{T,V} P) + (\partial_V C_V)^2	\Big]  \nonumber \\
& +C_V \Big[ (\partial_V P)^2+ T((\partial_V C_V)(\partial^2_V P)-T(\partial_{T,V} P^2))+2T(\partial_V P)(T(\partial_{T,T,V} P)-(\partial^2_V C_V))
\Big] \bigg \}.    \label{scalarCurva}
\end{align}
It should be noted that the sign of scalar curvature characterizes the type of interaction between two microscopic molecules in a given system \cite{ruppeiner2010thermodynamic}, i.e. $R>0$ and $R<0$ represent dominated repulsive interaction and attractive interaction, respectively. For VdW fluid, the study shows that it has a negative scalar curvature indicating the dominated attractive interaction among these underlying molecules. In addition, when $R=0$, the interaction of repulsion and attraction reaches equilibrium \cite{dolan2015intrinsic}.

\begin{figure}
\centering
\subfigure[]{\includegraphics[width=7cm]{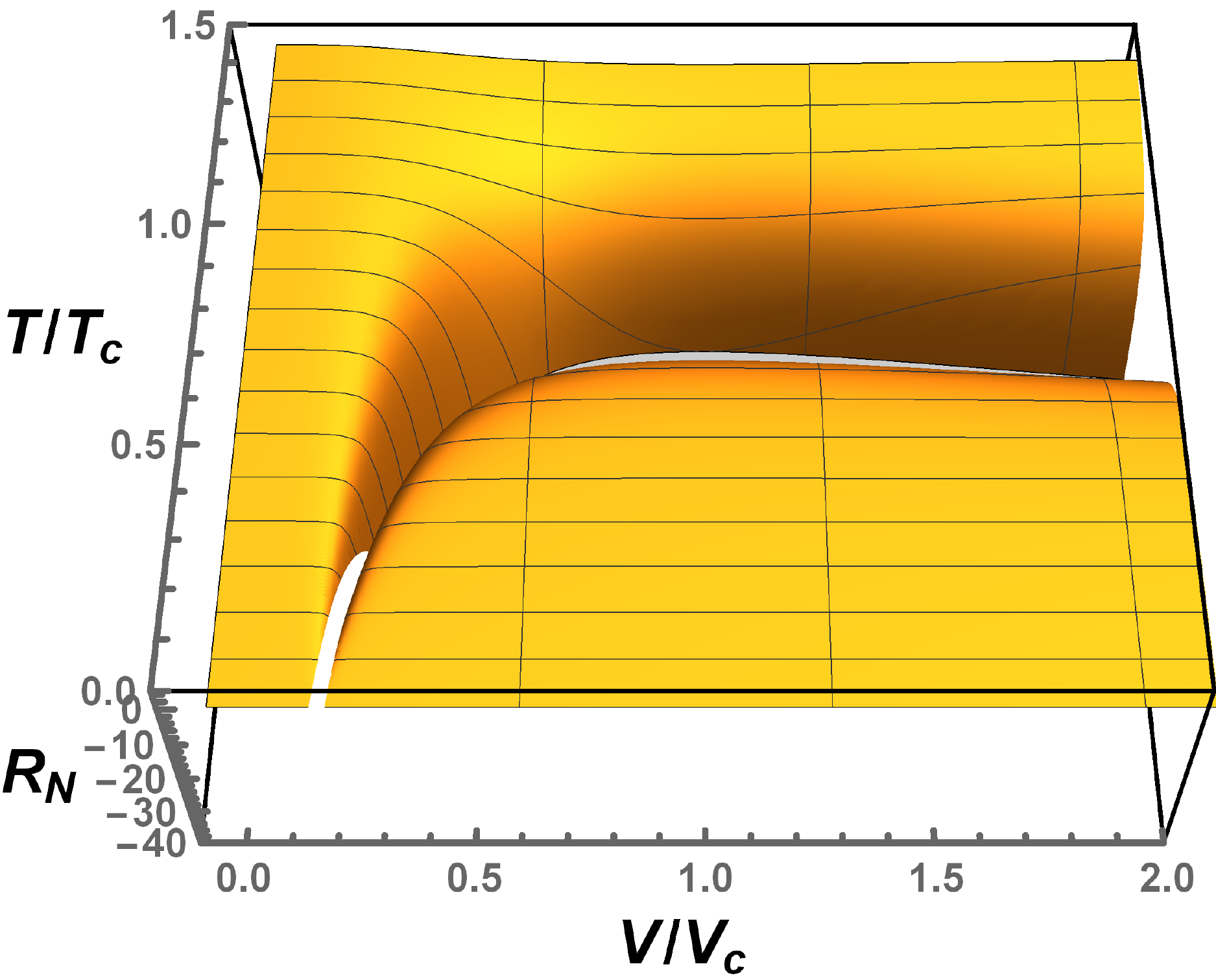} \label{fig:vdwrnvt_8a}}
\subfigure[]{\includegraphics[width=7cm]{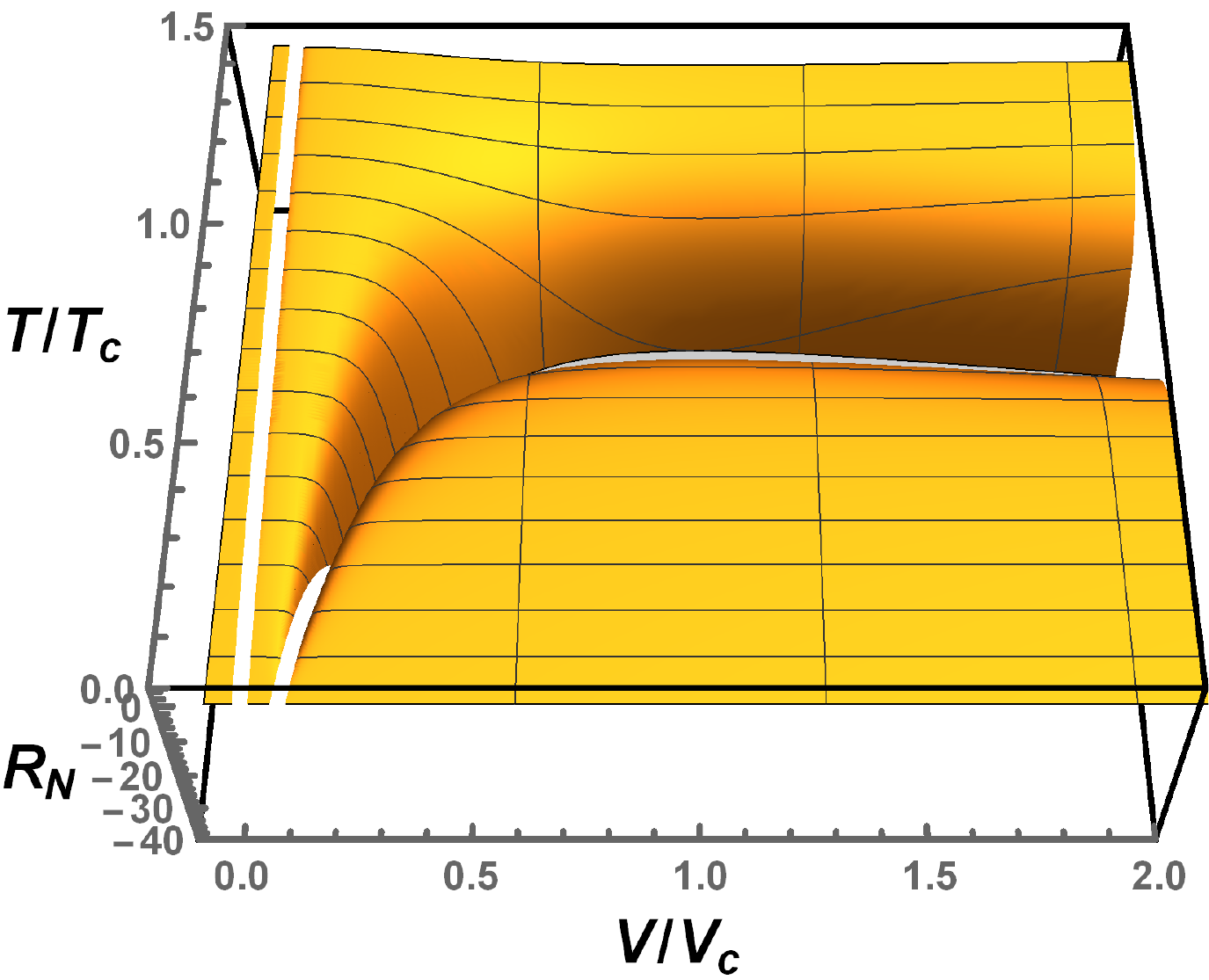}	\label{fig:rnvt_8b}}	
\caption{Behaviors of the normalized scalar curvature $ R_N $ as a function of $ \tilde{V} $ and $ \tilde{T} $ for the charged EH-AdS black hole. (a) VdW phase transition type case with $Q=1.0$ and $a=-1.5$. (b) Reentrant phase transition case with $Q=1.0$ and $a=1.0$.}
\label{fig:rnvt_8}
\end{figure}

For the charged EH-AdS black hole, the heat capacity at constant volume vanishes. We adopt the treatment of Ref. \cite{Wei:2019yvs}, the new normalized scalar curvature $R_N$ is
\begin{align}
R_{\rm N}&=R C_{V} \nonumber  \\
&=\frac{(\partial_{V}P)^{2}-T^{2}(\partial_{V,T}P)^{2}+2T^{2}(\partial_{V}P)(\partial_{V,T,T}P)}{2(\partial_{V}P)^{2}}.
\label{Rn}
\end{align}
Using the equation of state (\ref{state}), the normalized scalar curvature of the charged EH-AdS black hole reads
\begin{equation}\label{EHRn}
R_{\rm N}=\frac{A \left(A-2B\right)}{2\left( A-B \right)^2},
\end{equation}
where
\begin{align*}
A &= 16 \pi ^2 a Q^4-12 \sqrt[3]{6} \pi ^{2/3} Q^2 V^{4/3}+9 V^2,\\
B &= 9 \sqrt[3]{6} \pi ^{2/3} T V^{7/3}.
\end{align*}
We plotted the scalar curvature $R_N$ as a function of $ \tilde{V} $and $ \tilde{T} $ in Fig. \ref{fig:rnvt_8} for the small/large black hole phase transition and reentrant phase transition. It can be observed that the surface of the normalized scalar curvature is concave where the scalar curvature diverges. Besides, compared with the VdW type phase transition, a new divergence occurs in the reentrant phase transition when the reduced volume $\tilde{V}$ is close to zero.

\subsection{Van der Waals type phase transition with $a<0$}

As we shown above, there is the small/large black hole phase transition of VdW-like for the case.

\begin{figure}
\begin{center}
	\subfigure[]{\label{fig:VdWRnva_9a}
		\includegraphics[width=7cm]{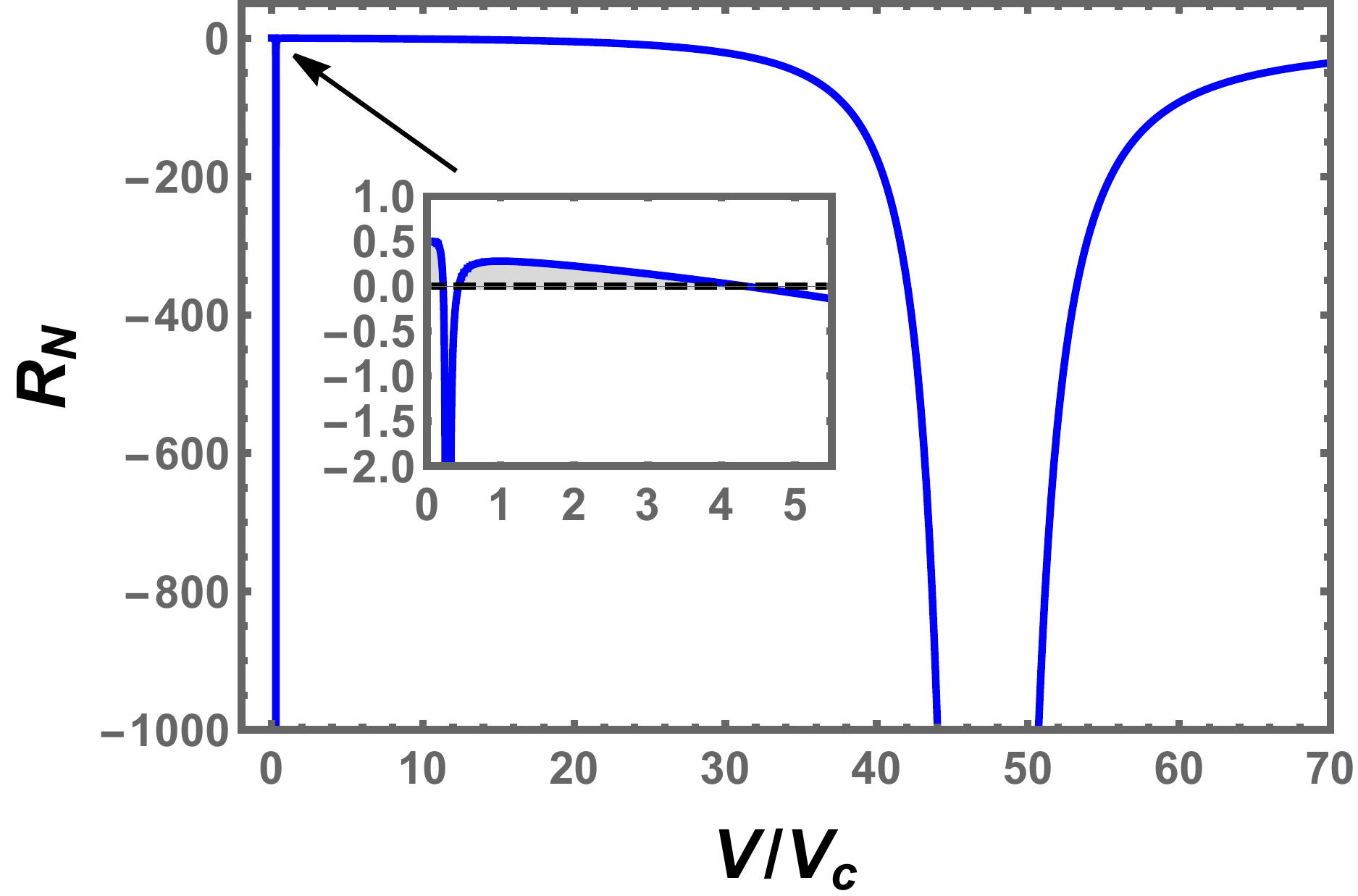}}
	\subfigure[]{\label{fig:VdWRnvb_9b}
		\includegraphics[width=7cm]{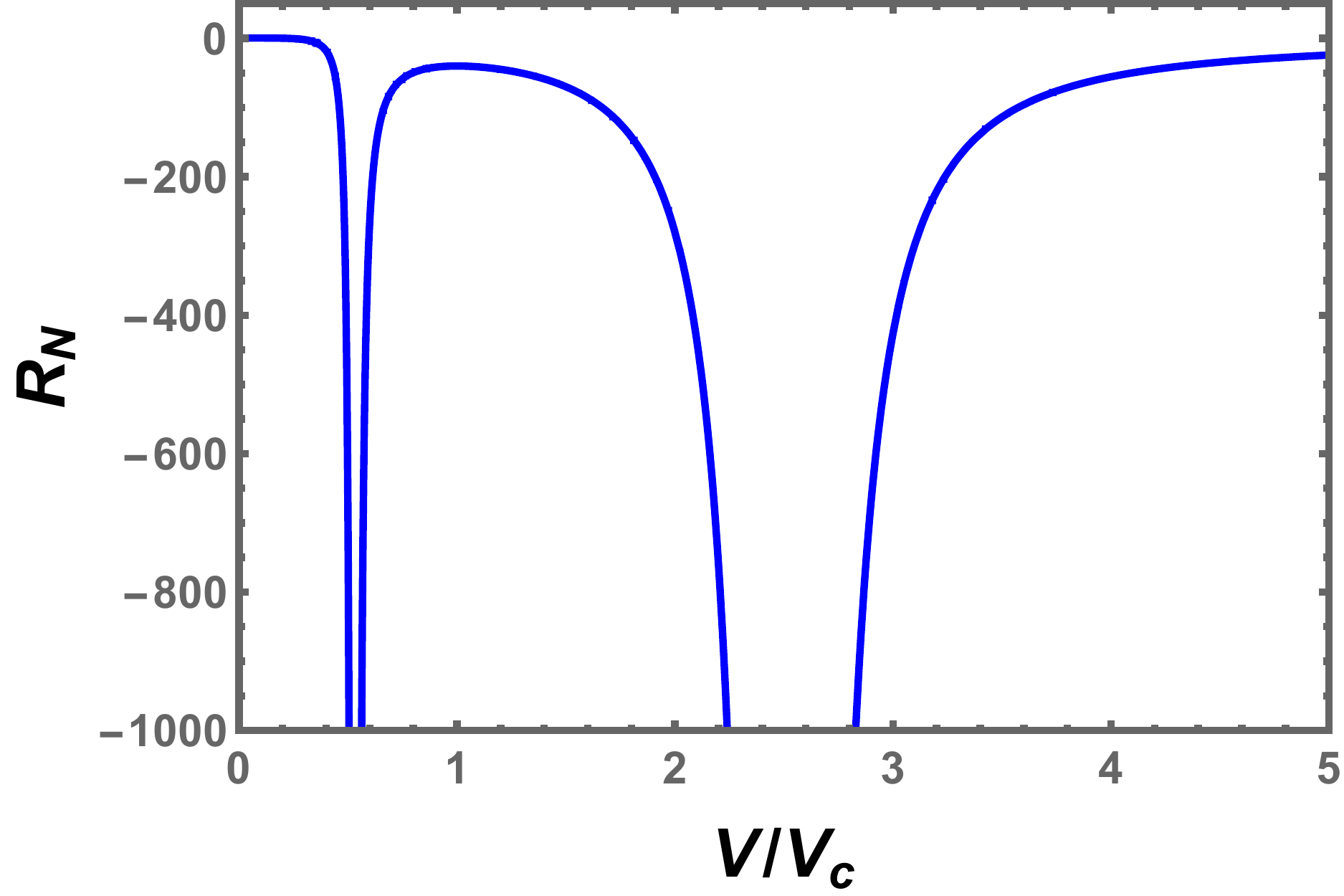}}\\
	\subfigure[]{\label{fig:VdWRnvc_9c}
		\includegraphics[width=7cm]{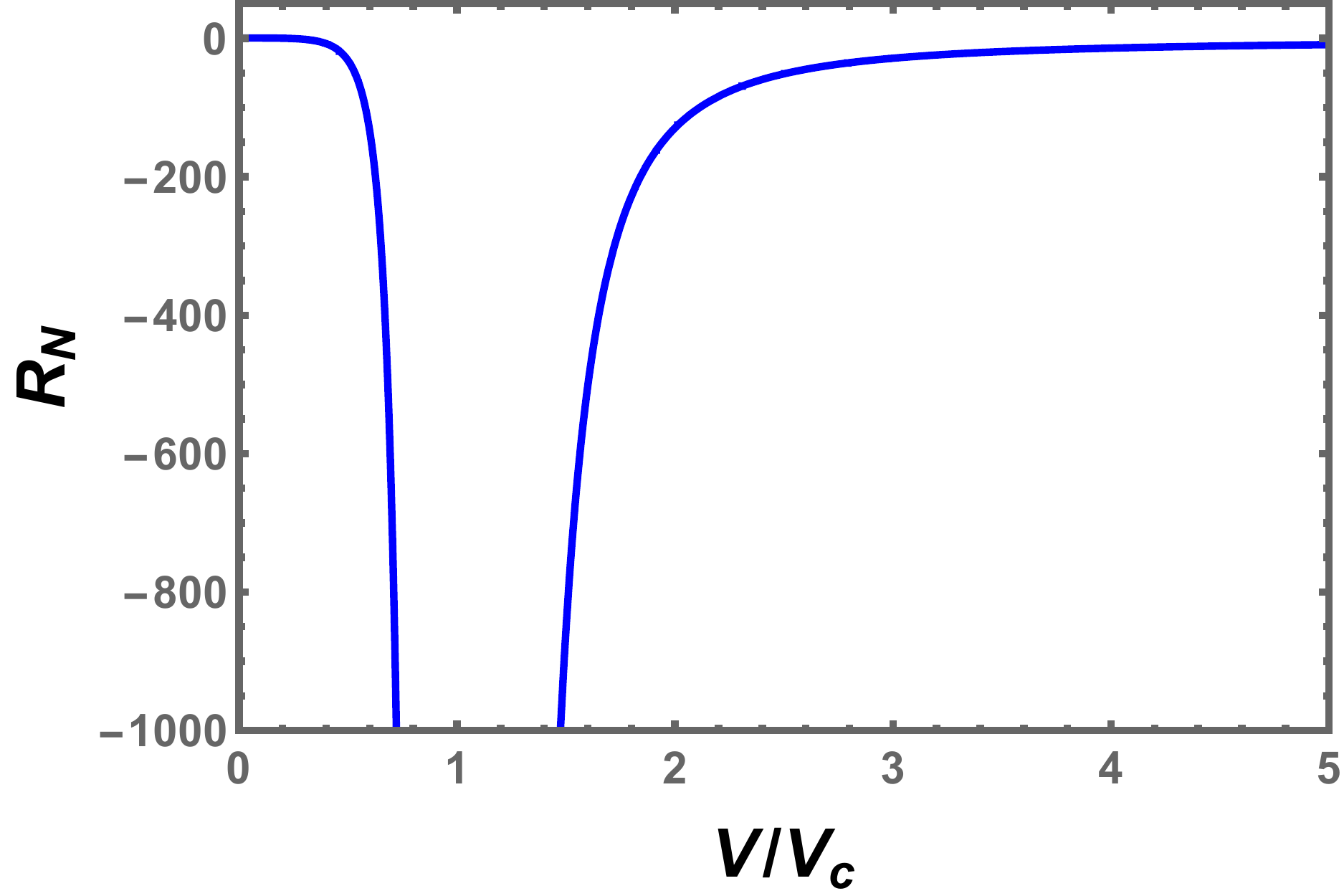}}
	\subfigure[]{\label{fig:VdWRnvd_9d}
		\includegraphics[width=7cm]{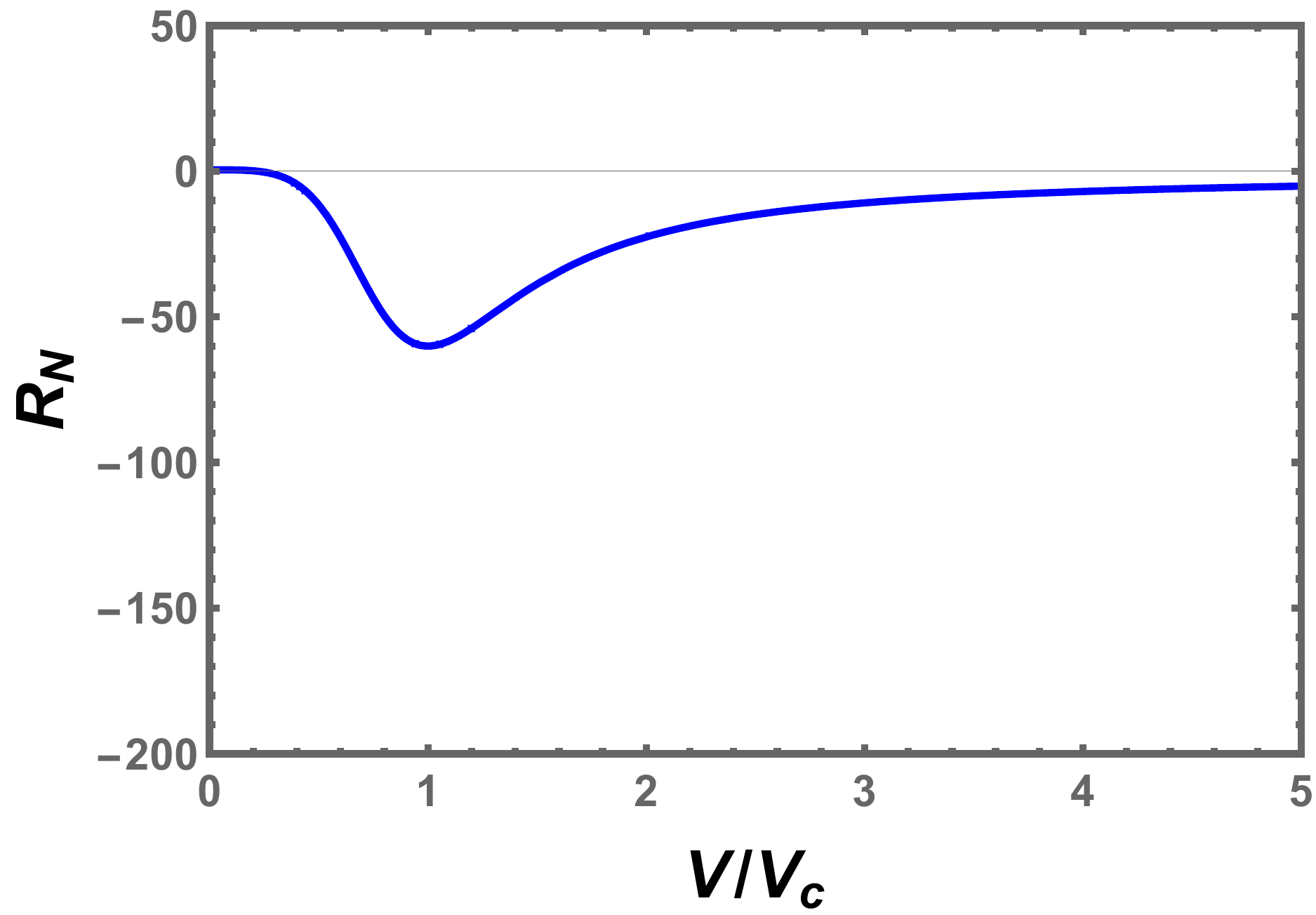}}\
\end{center}
\caption{The behaviors of the normalized scalar curvature $R_N$ with the reduced volume $\tilde{V}$ at a constant temperature. (a) $ \tilde{T}=0.4$. (b) $\tilde{T}=0.8$. (c) $\tilde{T}=1.0$. (d) $\tilde{T}=1.2$. The inset shows the enlarged portion near the origin and $R_N$ has positive values, which is depicted by shadow regions. We have set $Q=1.0$ and $a=-1.5$.}
\label{fig:VdWrnv_9}
\end{figure}

In order to show the details, we take $Q=1.0$ and $a=-1.5$, and illustrate the normalized scalar curvature $R_{N}$ as a function of $\tilde{V}$ for fixed reduce temperature $\tilde{T}$= 0.4, 0.8, 1.0 and 1.2 in Fig. \ref{fig:VdWrnv_9}.
For $T<T_c$, there are two divergent points of the normalized scalar curvature $R_N$. It can be found that these two points will come closer as the temperature increases. When the critical temperature is approached, these two points merge. When $T>T_c$, the normalized scalar curvature $R_N$ does not diverge while keeps finite values. In most of the parameter space, $R_N$ is negative, while in a small region of $\tilde{V}$ for $\tilde{T}$= 0.4 shown in the inset of Fig. \ref{fig:VdWRnva_9a}, $R_N$ is positive. This phenomenon indicates that the dominated repulsive interaction may exist in some parameter regions.

From the analysis of the normalized scalar curvature $ R_{N} $, it is easy to know that the divergent point of $R_{N}$ satisfies the spinodal point condition $(\partial_{V} P)_{T}=0$, which gives
\begin{equation}\label{Tsp}
T_{sp}=-\frac{-16 \pi ^2 a Q^4+12 \sqrt[3]{6} \pi ^{2/3} Q^2 V^{4/3}-9 V^2}{9 \sqrt[3]{6} \pi ^{2/3} V^{7/3}}.
\end{equation}
After a simple calculation, the specific expression of the sign changing curve corresponding to $R_{N}=0$ reads
\begin{equation}\label{Tsc}
T_{sc}= \frac{T_{sp}}{2}=-\frac{-16 \pi ^2 a Q^4+12 \sqrt[3]{6} \pi ^{2/3} Q^2 V^{4/3}-9 V^2}{18 \sqrt[3]{6} \pi ^{2/3} V^{7/3}},
\end{equation}
which indicates that the temperature of the sign changing curve is half of the temperature of the spinodal curve \cite{Wei:2019uqg,Wei:2019yvs}.

\begin{figure}[h]
\centering
\subfigure[]{\includegraphics[width=7cm]{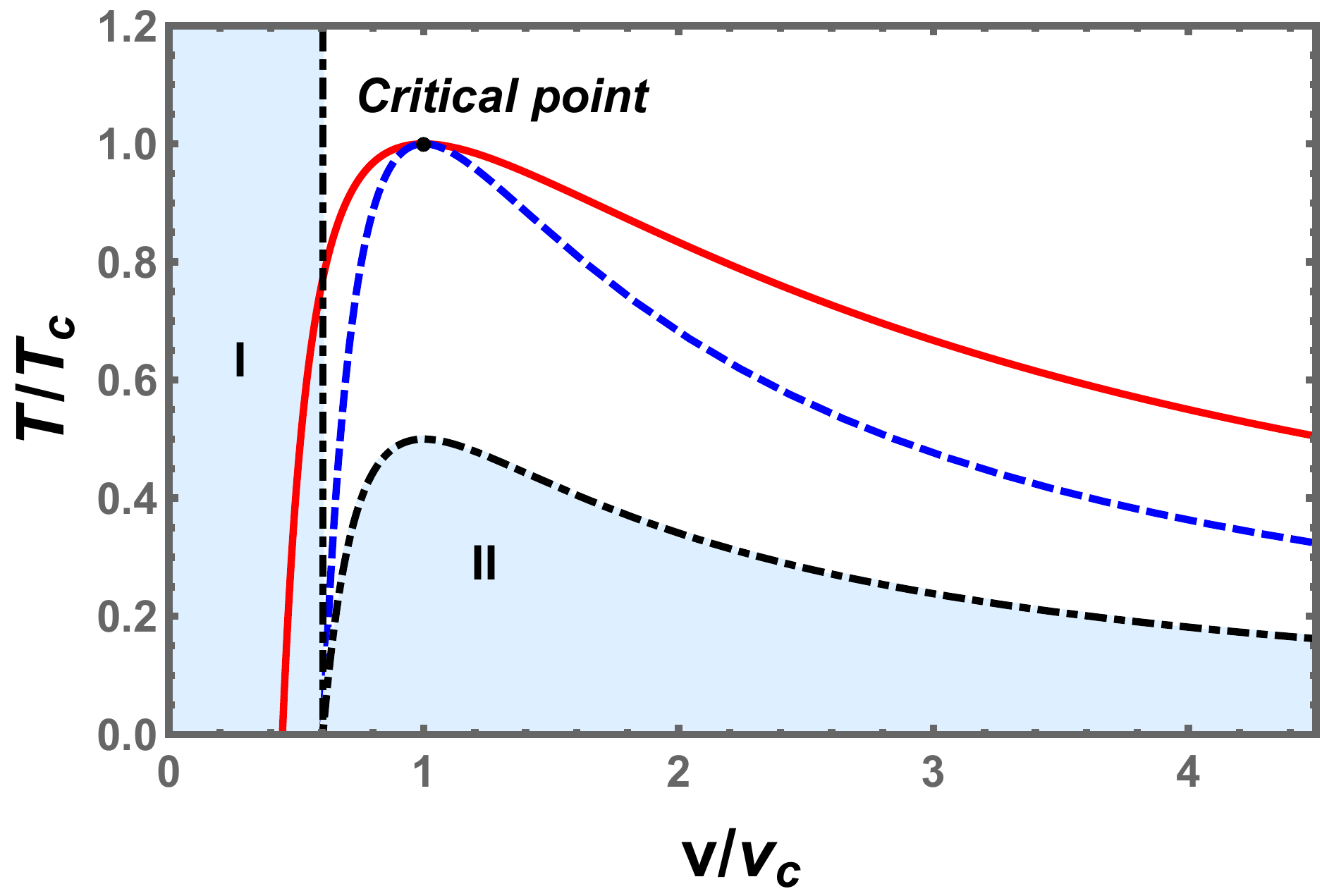}\label{fig:vdwcoetsptsc_10a}}
\subfigure[]{\includegraphics[width=7cm,height=4.6cm]{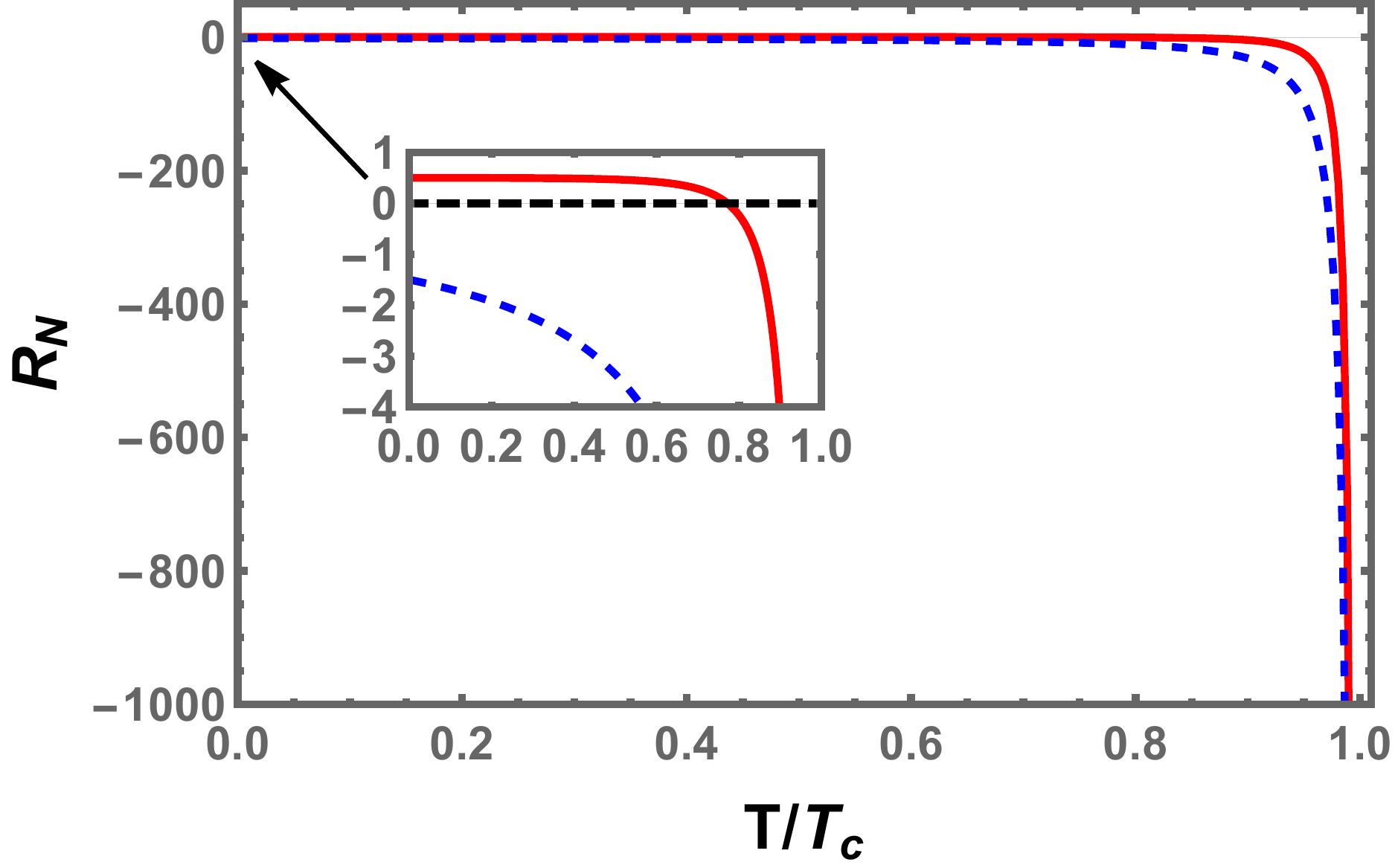}\label{fig:VdWrnt_10b}}
\caption{(a) The sign changing curve of $R_N$ (black dot dashed line), spinodal curve (blue dashed line) and the coexistence curve (red solid line) for VdW type phase transition case. The shadow region marked with \textbf{I} and \textbf{II} corresponds to positive $R_N$, otherwise $R_N$ is negative. (b) The behavior of normalized Ruppeiner curvature scalar $R_N$ along the coexistence curve. The red solid line and blue dashed line correspond to SBH and LBH, respectively. The change in nature of interaction of SBH is shown in inset. We have set $Q=1.0$ and $a=-1.5$.}
\end{figure}

In Fig. \ref{fig:vdwcoetsptsc_10a}, the sign changing curve (black dotdashed), spinodal curve (blue dashed), and the coexistence curve (red solid) are displayed. The shadow regions I and II have positive scalar curvature, while the other regions are for negative scalar curvature. Considering that the equation of state for the charged EH-AdS black hole in region II is no longer valid, this region needs to be excluded. However, in region I, the small black hole with high temperature still admits the dominated repulsive interaction. This is similar to the situation of the charged RN-AdS black holes, while different from the neutral black hole in Gauss-Bonnet gravity \cite{Wei:2019ctz,Zhou:2020vzf}.

We show the behavior of $R_N$ along the coexistence small and large black holes in Fig. \ref{fig:VdWrnt_10b}. It is easy to see that both them go to negative infinity at the critical temperature indicating the existence of the critical exponent. Moreover, we can also find that $R_N$ of the coexistence small black hole is always above that of the large black hole. At low temperature, the coexistence small black hole has a positive value, which is consistent with that of Fig. \ref{fig:vdwcoetsptsc_10a}.

Since the critical exponent can provide us with some universal properties, we now turn to calculate it for the scalar curvature at the critical point. Here it is natural to assume the normalized scalar curvature $R_N$ near the critical point has the following form \cite{johnston2014advances}
\begin{equation}
R_{N}\sim-(1-\tilde{T})^{-\alpha},
\end{equation}
or equivalently,
\begin{equation}
\ln|R_{N}|=-\alpha\ln(1-\tilde{T})+\beta.\label{FitR}
\end{equation}
Via calculating $R_N$ near the critical point, we obtain the following fitting results
\begin{eqnarray}
\ln|R_{N}|&=&-2.0365\ln(1-\tilde{T})-2.4466, \quad \text{for coexistence SBH,}\label{fitsmall}\\
\ln|R_{N}|&=&-1.9636\ln(1-\tilde{T})-1.7135, \quad \text{for coexistence LBH.}\label{fitlarge}
\end{eqnarray}

\begin{figure}[h]
\centering
\subfigure[]{
	\includegraphics[width=7cm]{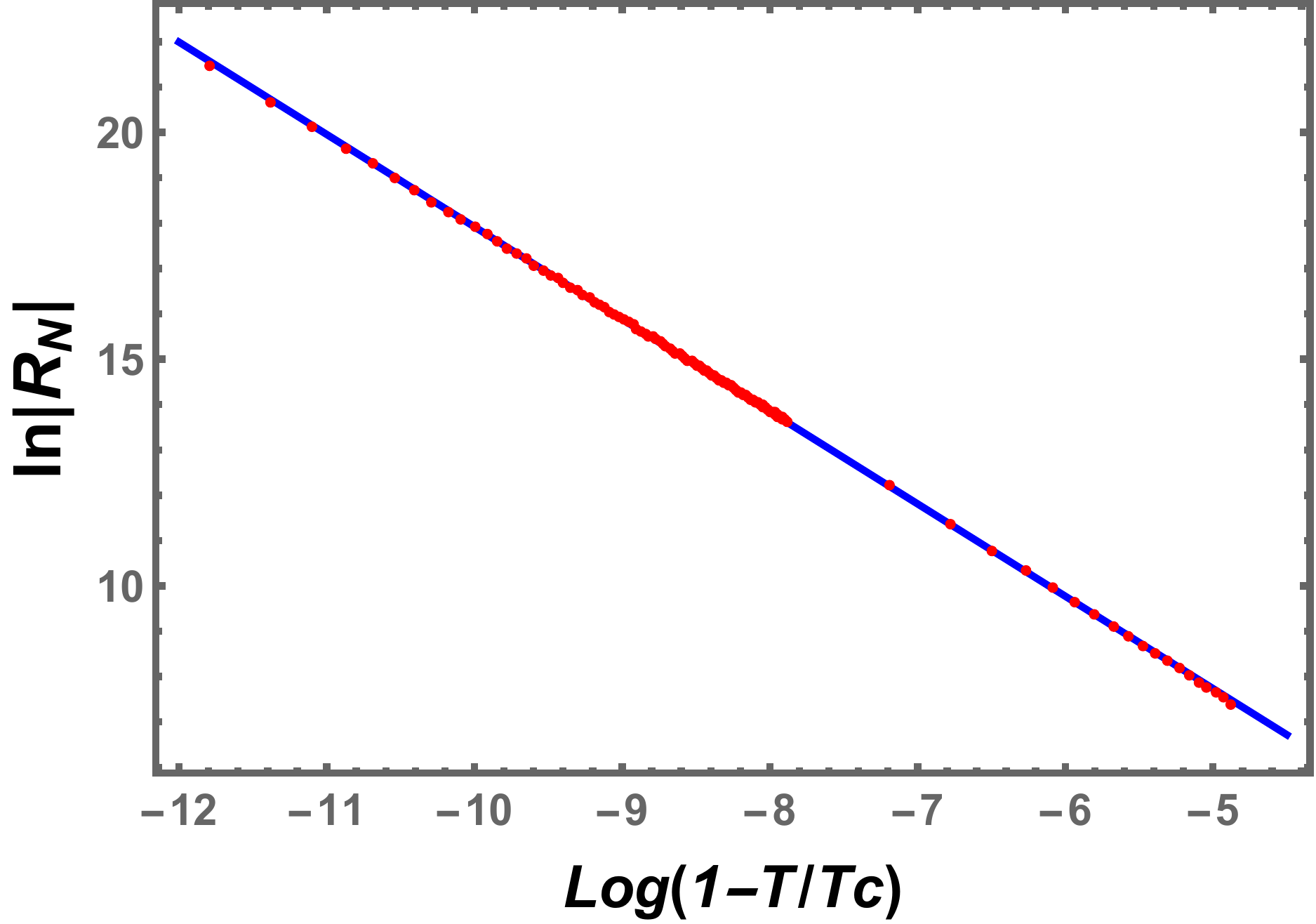}
	\label{fig:vdwfitsamll_11a}}
\subfigure[]{
	\includegraphics[width=7cm]{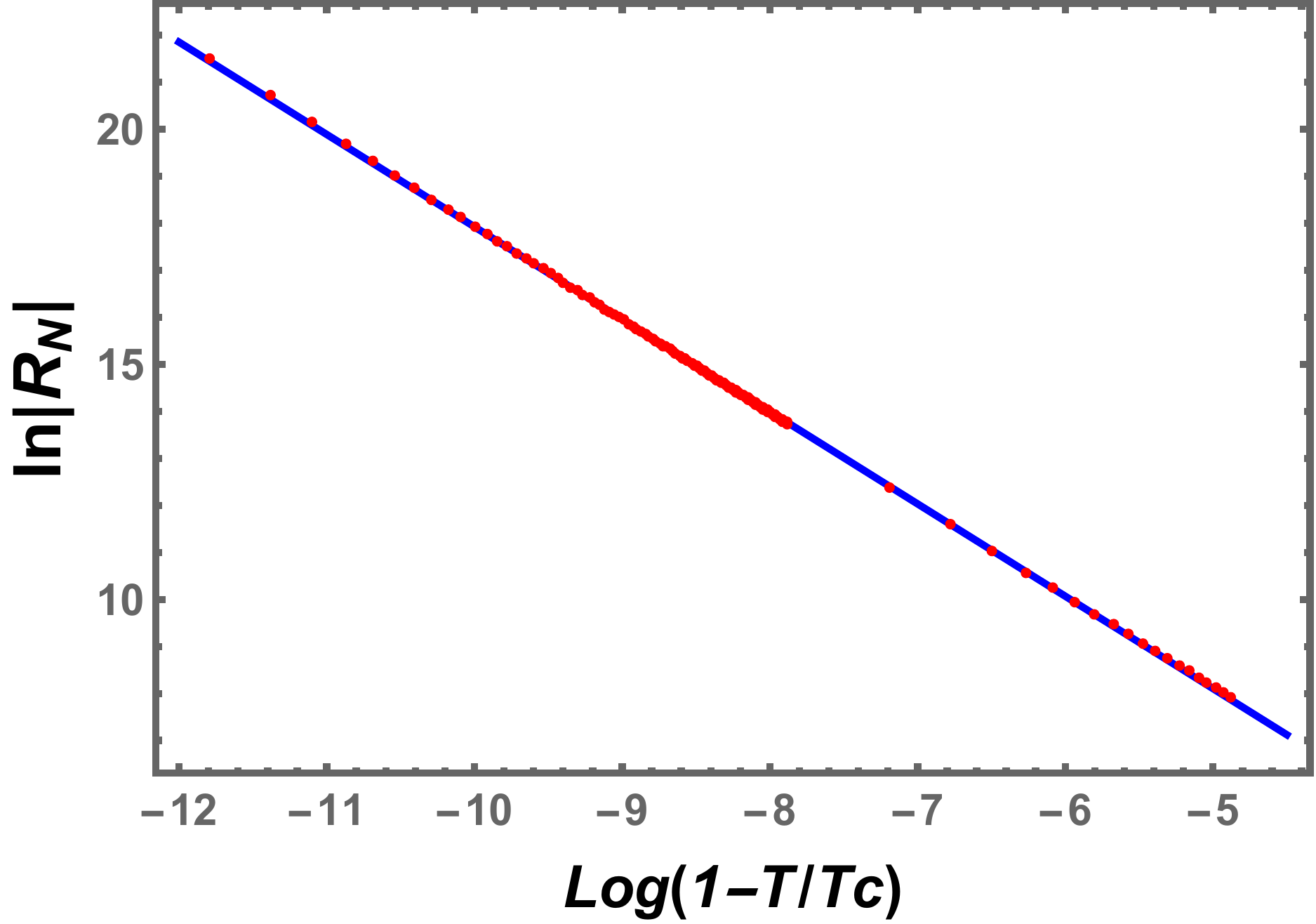}
	\label{fig:vdwfitlarge_11b}}
\caption{The fitting curves of $\ln |R_N|$ vs. $\ln (1-\tilde{T})$ near the critical point. The red dots are numerical data and blue solid lines are obtained from the fitting formulas. (a) The coexistence SBH branch. (b) The coexistence LBH branch.}
\label{fig:vdwfitline_11}
\end{figure}

The numerical results (small red dots) and fitting results (blue solid lines) are shown in Fig. \ref{fig:vdwfitline_11}. It is obvious that the numerical and fitting results are highly consistent with each other. The slopes obtained by fitting the coexistence curves of small and large black holes are respectively $\alpha_{SBH}$=2.0365 and $\alpha_{LBH}$=1.9636. Taking into account the error of the numerical calculation, we can say that the critical exponent is $ \alpha=2 $, which is the same as that of the VdW fluid.

Moreover, using the intercept $ \beta $ obtained by fitting results, we obtain a dimensionless constant
\begin{equation}\label{beta}
R_{N}(1-\tilde{T})^{2}=e^{(-2.4466-1.7135)/2}= -0.124924\simeq-\frac{1}{8}.
\end{equation}
It is the same value with the charged AdS black holes and VdW fluids, and is also consistent with the result obtained in Ref. \cite{Magos:2020ykt} via expanding the equation of state near the critical point at the first leading term.

\subsection{Reentrant phase transition case with $0\leq a\leq \frac{32 Q^2}{7}$}

\begin{figure}
\begin{center}
	\subfigure[]{\label{fig:Rnva_12a}
		\includegraphics[width=7cm]{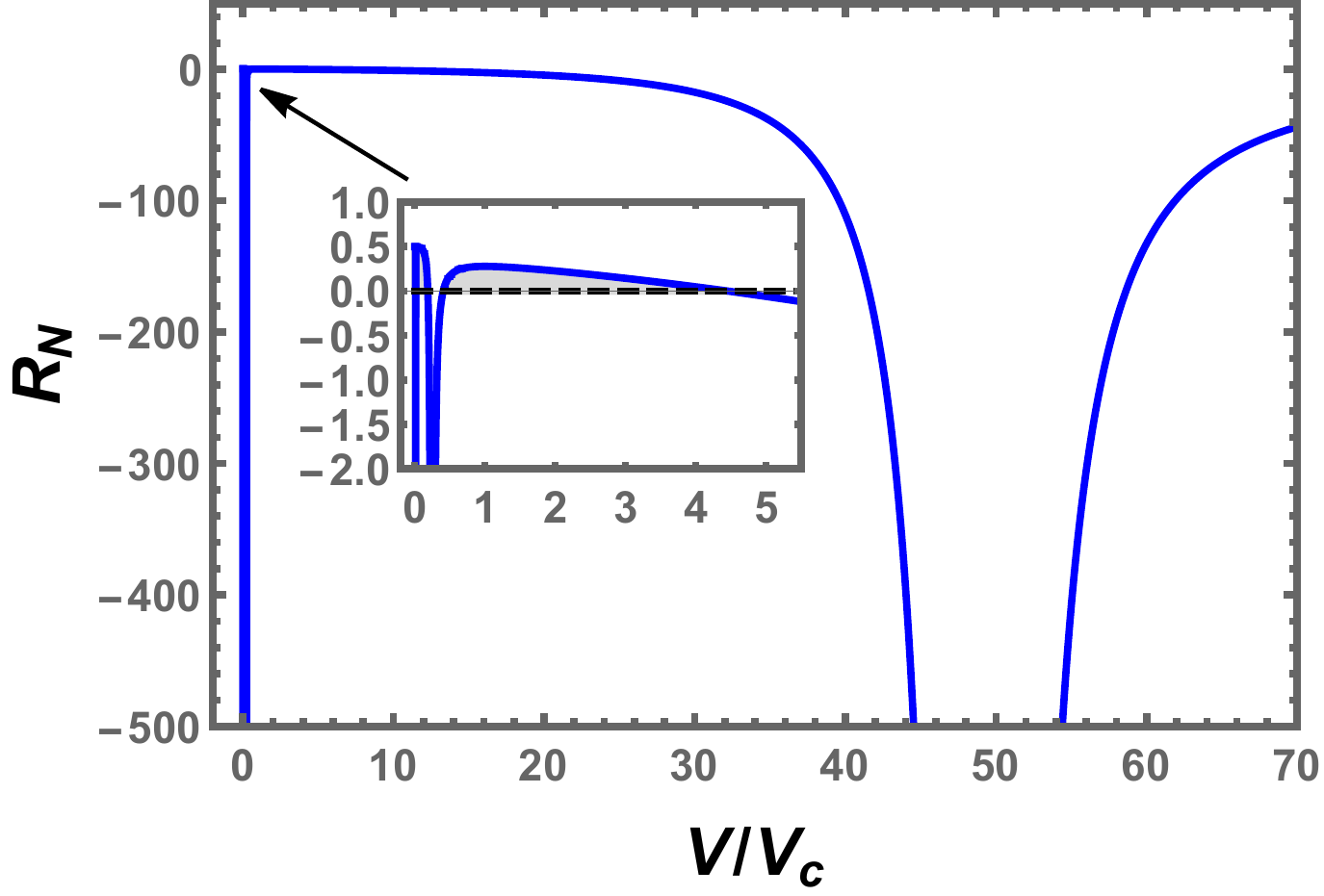}}
	\subfigure[]{\label{fig:Rnvb_12b}
		\includegraphics[width=7cm]{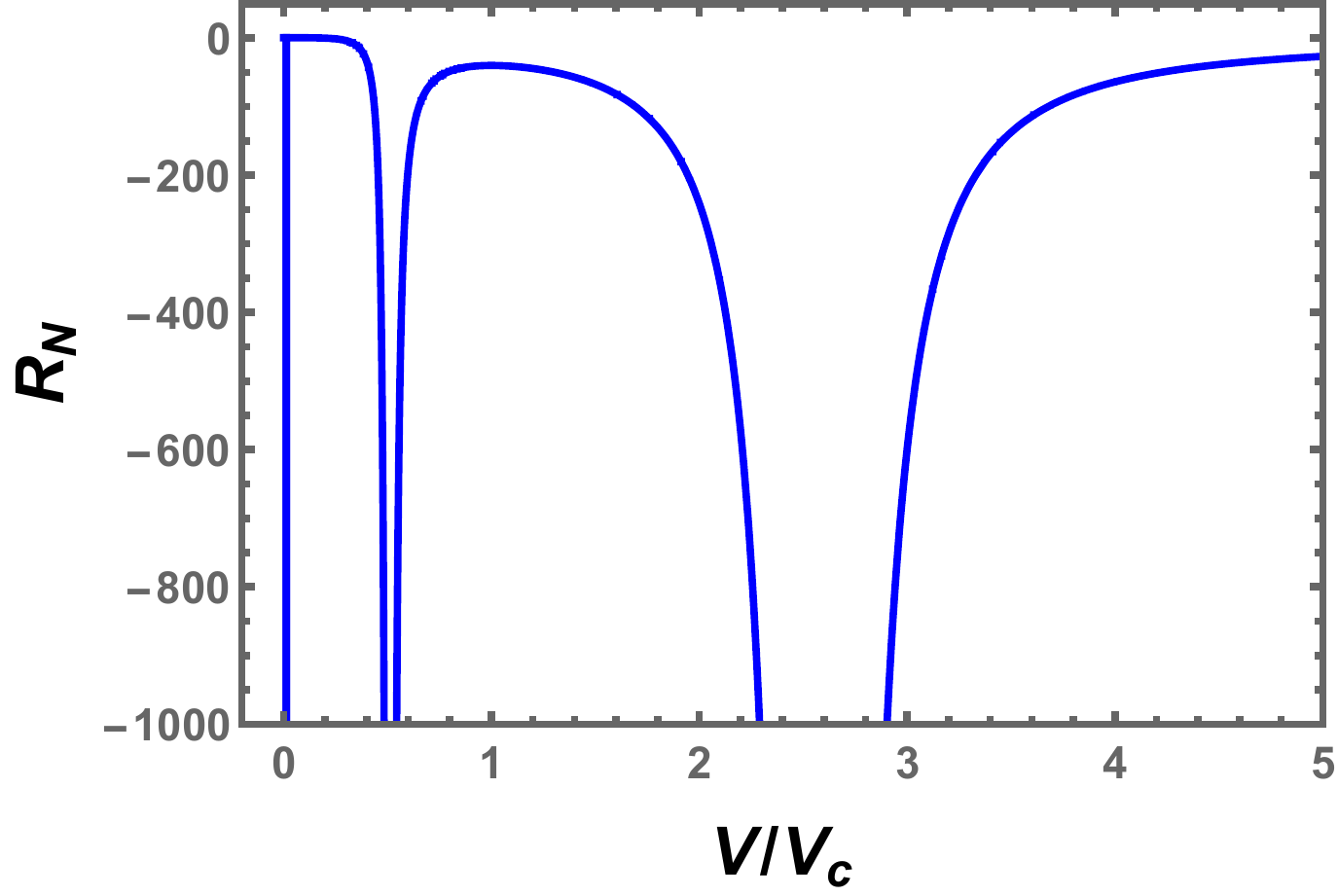}}\\
	\subfigure[]{\label{fig:Rnvc_12c}
		\includegraphics[width=7cm]{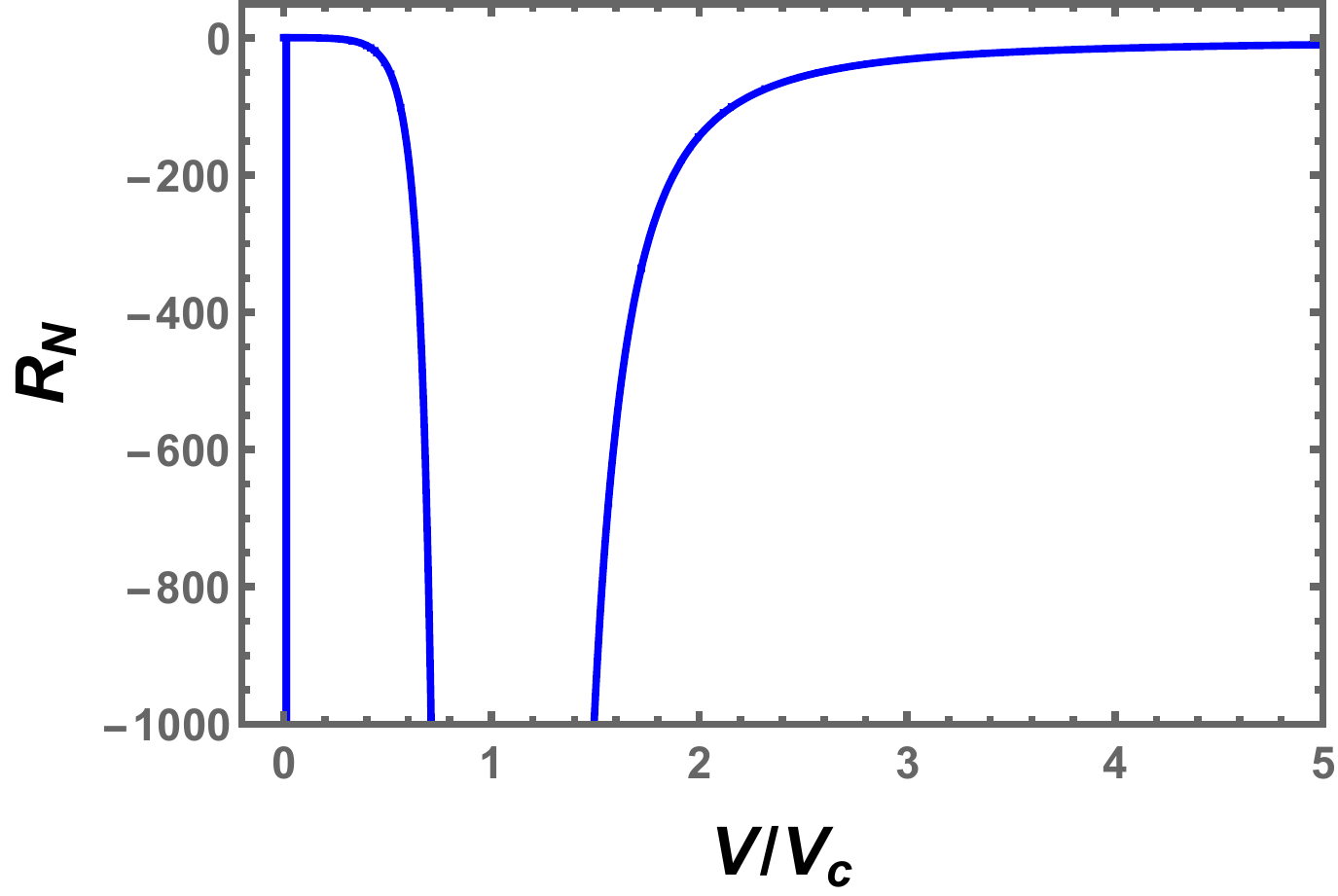}}
	\subfigure[]{\label{fig:Rnvd_12d}
		\includegraphics[width=7cm]{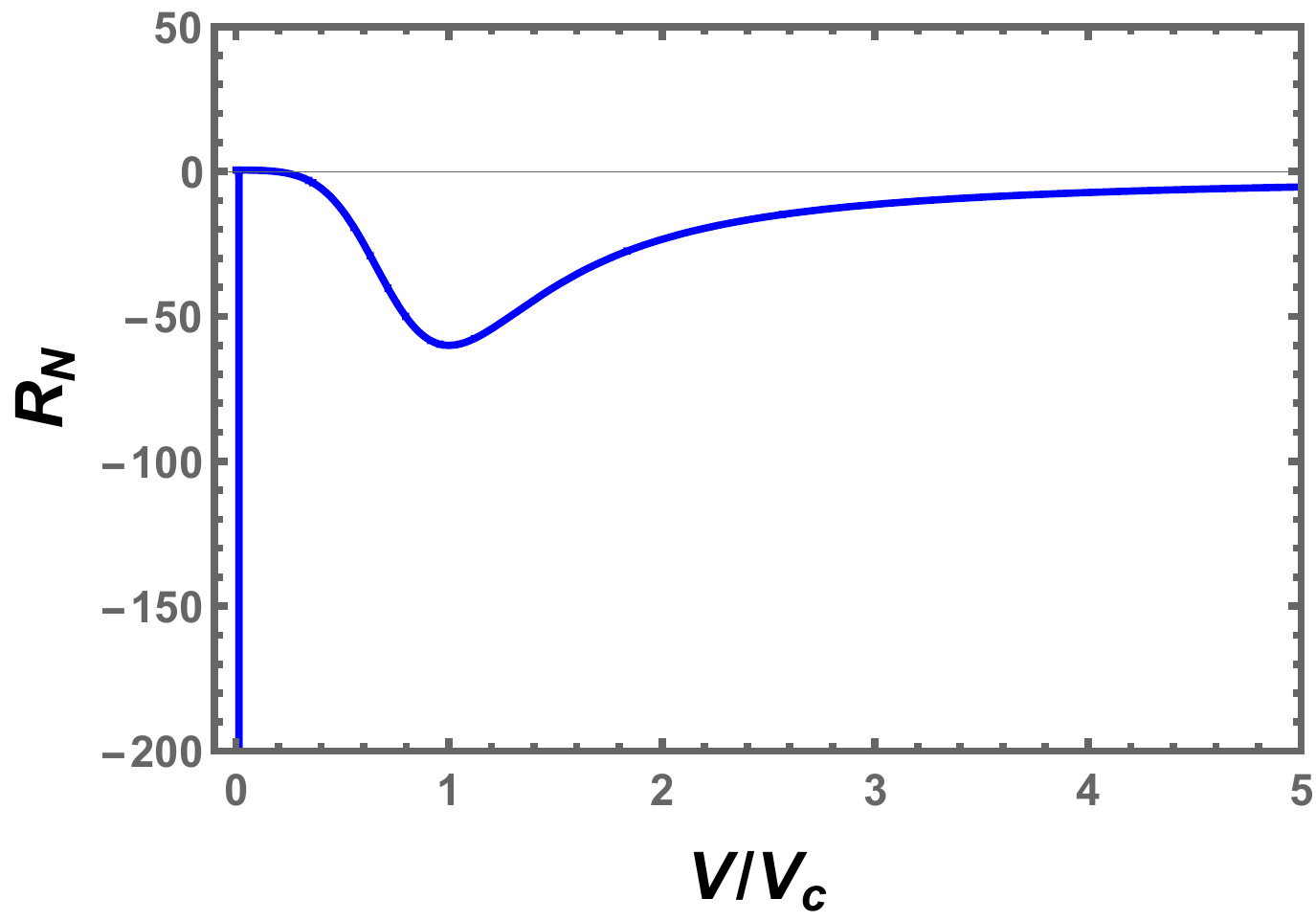}}\
\end{center}
\caption{The behaviors of the normalized scalar curvature $R_N$ with the reduced volume $\tilde{V}$ at a constant temperature for reentrant phase transition case. (a) $\tilde{T}=0.4$. (b) $\tilde{T}=0.8$. (c) $\tilde{T}=1.0$. (d) $\tilde{T}=1.2$. The inset shows the enlarged portion near the origin and $ R_N $ has positive values. We have set $Q=1.0$ and $a=1.0$.}
\label{fig:rnv_12}
\end{figure}

We explore the underlying microstructure of the black hole that results in reentrant phase transition with $Q$=1 and $a$=1 by using the similar method. The behavior of $R_N$ against the reduced thermodynamic volume $ \tilde{V} $ for a fixed temperature is shown in Fig. \ref{fig:rnv_12}. Comparing with the VdW phase transition, we observe that the scalar curvature of the reentrant phase transition has an additional divergent point near the origin. The reason is when $T<T_c$, there are three extremal points on each isothermal curve, see Fig. \ref{fig:pv_1b}. As the temperature increases, the evolution behavior of other two divergent points is very similar to the case of the VdW-like phase transition. Furthermore, when the temperature is higher than its critical value, the additional divergent point still exists. On the other hand, when the reduced volume is close to the origin, one observes a positive scalar curvature as expected.

\begin{figure}
\centering
\subfigure[]{\includegraphics[width=7cm]{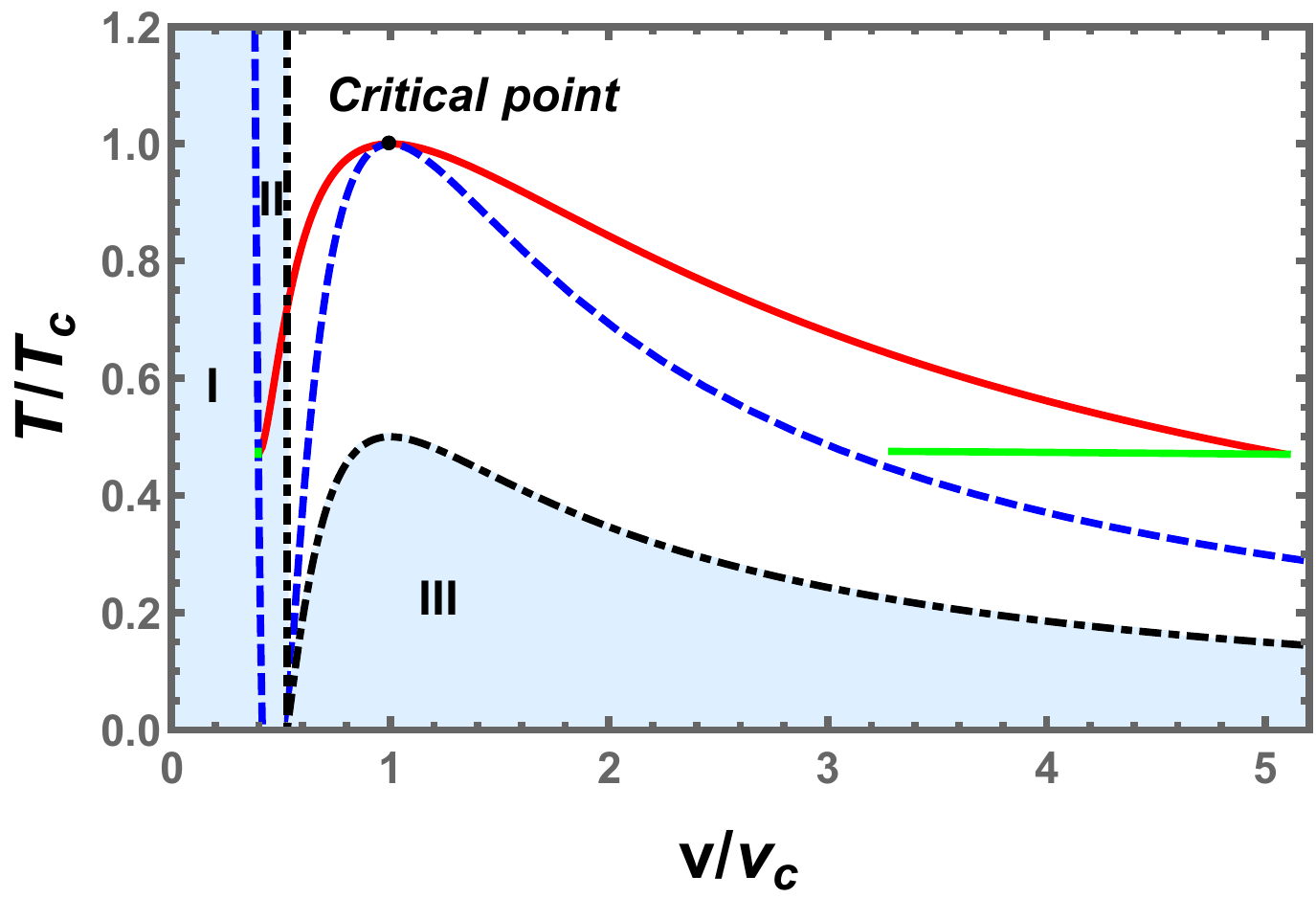}\label{fig:coetsptsc_13a}}
\subfigure[]{\includegraphics[width=7cm,height=4.6cm]{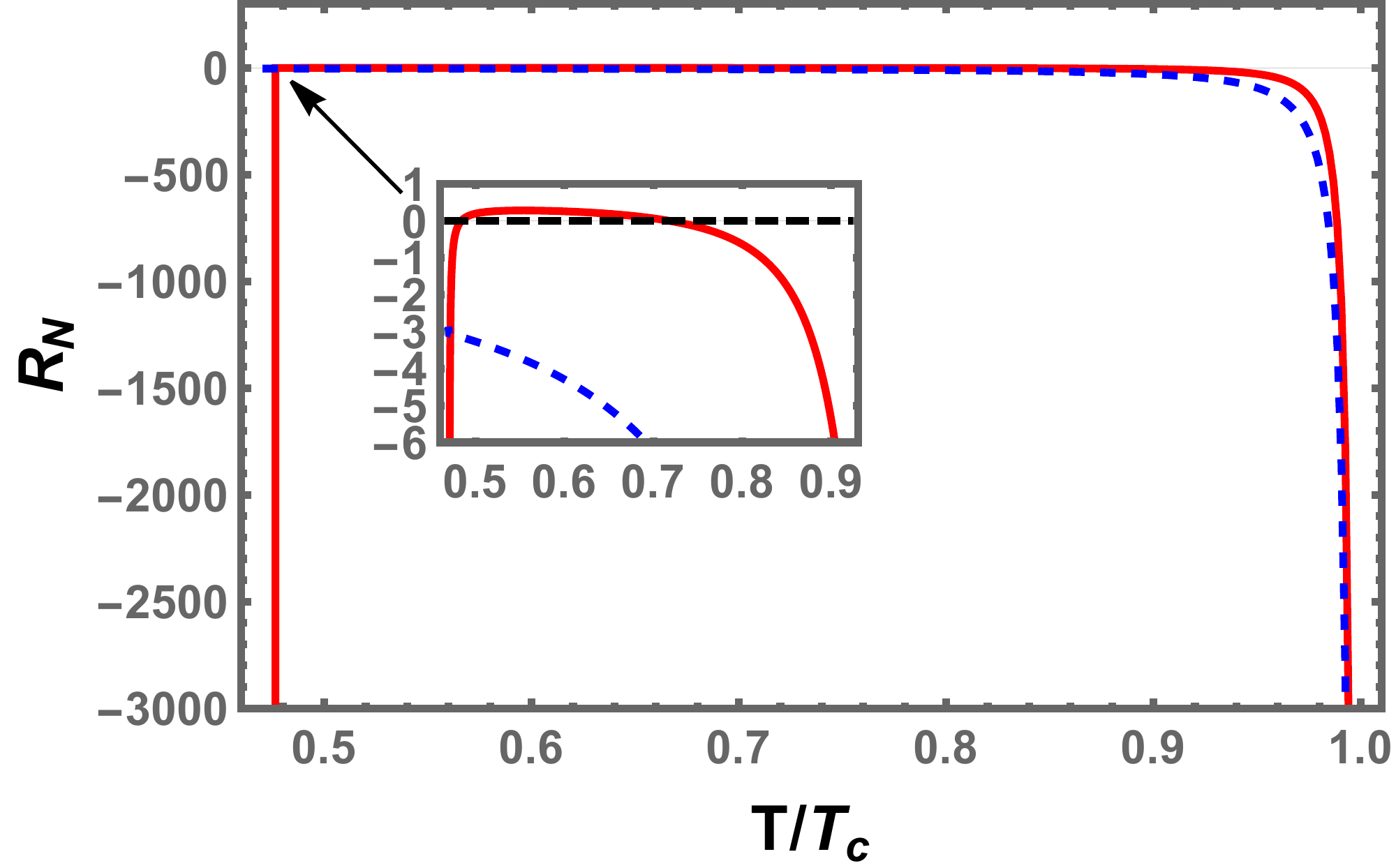}\label{fig:rnt_13b}}
\caption{(a) The sign changing curve of $ R_N $ (black dot dashed line), spinodal curve (blue dashed line), first-order coexistence curve (red solid line) and zeroth-order phase transition line (green solid line) for reentrant phase transition case. The shadow region marked with \textbf{I}, \textbf{II} and \textbf{III} corresponds to positive $ R_N $, otherwise $ R_N $ is negative. (b) The behavior of normalized scalar curvature $ R_N $ along the coexistence curve. The red (solid) line and blue (dashed) line correspond to SBH and LBH, respectively. The change in sign of $ R_N $ of the SBH is shown in inset.  }
\end{figure}

The first-order coexistence curve, zeroth-order phase transition curve, spinodal curve, and sign changing curve are depicted in Fig. \ref{fig:coetsptsc_13a}. In the shadow regions I, II, and III, the scalar curvature takes positive values. Compared with the VdW-like phase transition, the number of the shadow regions is three instead of two. It can be seen that the leftmost of the coexistence curve of small black hole ends at the spinodal curve and it indicates that there is a divergence. On the other hand, one should note that the black hole does not exist in region I.

We also plot the normalized scalar curvature $R_{N}$ along the coexistence small and large black hole curves as a function of temperature in Fig. \ref{fig:rnt_13b}. Along the coexistence curve, $R_N$ negatively diverges at the critical point for both the small and large black holes. However, different from the previous VdW-like phase transition, the scalar curvature of the small black hole gains a new divergence when the reduce temperature is around 0.48. While for the coexistence large black hole, there is no such phenomenon. The reason for this new divergence is that the leftmost of the coexistence small black hole curve is located on the spinodal curve, but the large black hole is not.

Near the critical point, the fitting results for the coexistence small and large black holes are given by
\begin{eqnarray}
\ln|R_{N}|&=&-2.0426\ln(1-\tilde{T})-2.5918, \quad \text{for coexistence SBH.}\label{fitsmall}\\
\ln|R_{N}|&=&-1.9536\ln(1-\tilde{T})-1.5253, \quad \text{for coexistence LBH.}\label{fitlarge}
\end{eqnarray}

\begin{figure}
\centering
\subfigure[Small BH]{
	\includegraphics[width=7cm]{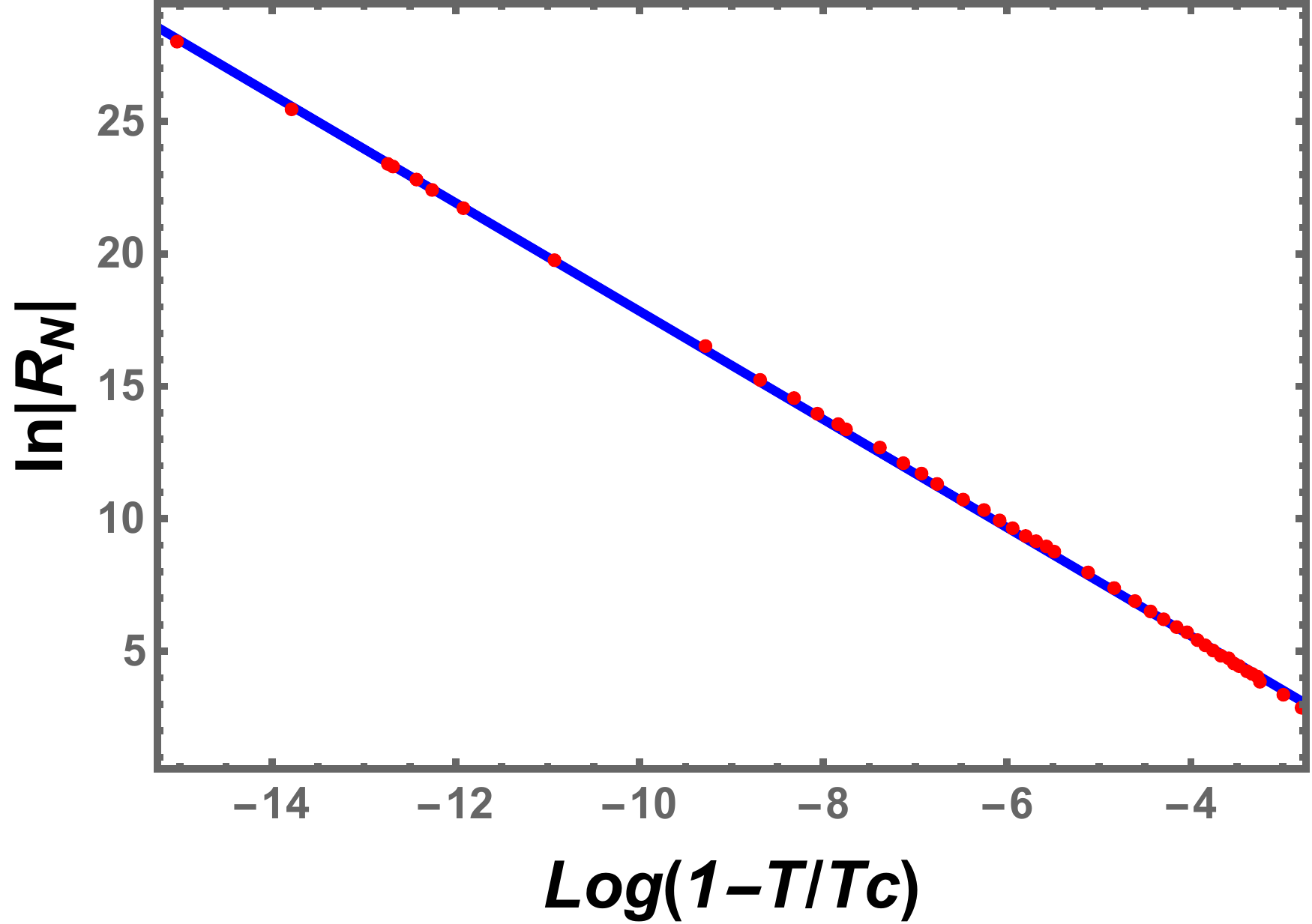}
	\label{fig:fitsamll_14a}}
\subfigure[Large BH]{
	\includegraphics[width=7cm]{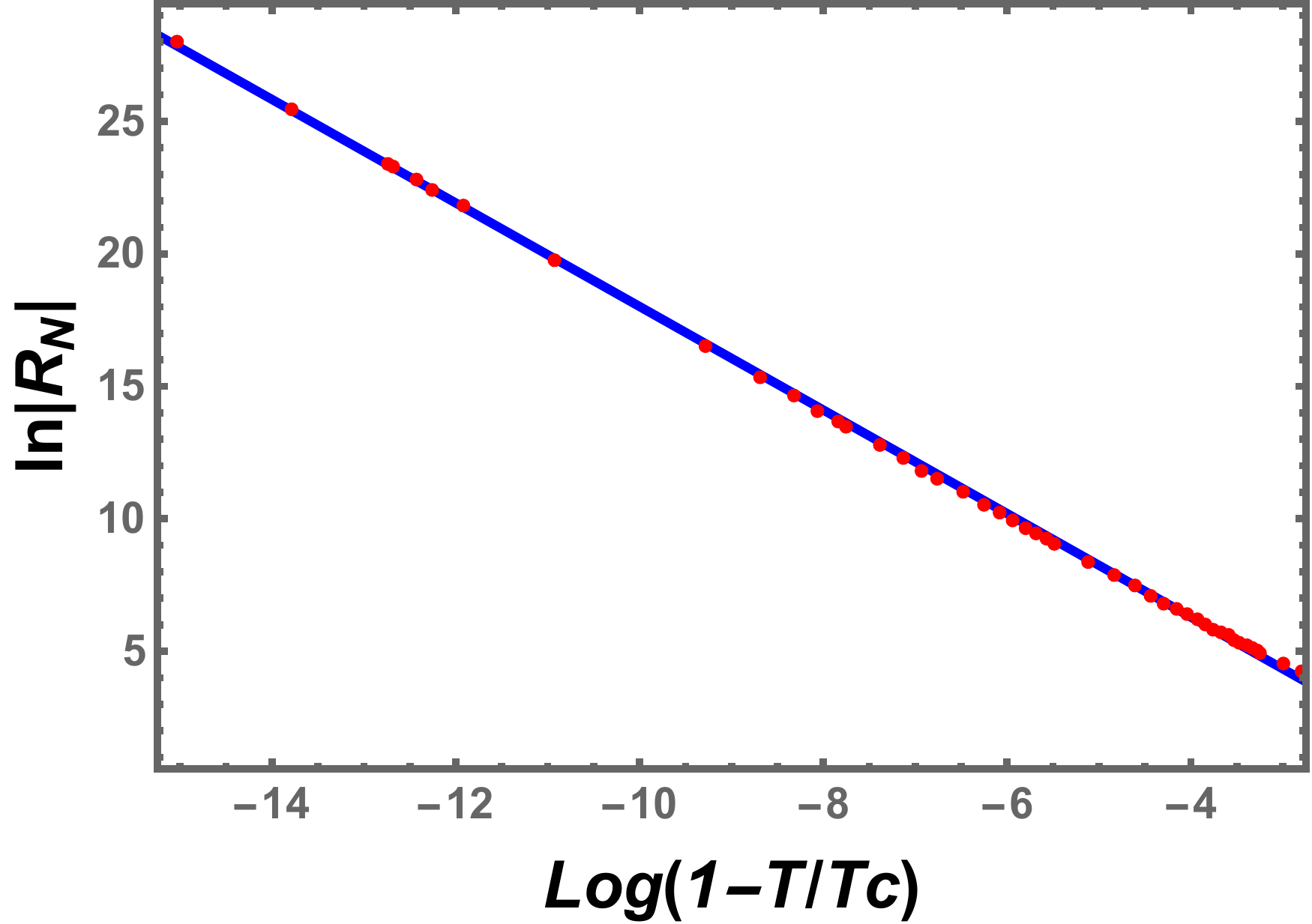}
	\label{fig:fitlarge_14b}}
\caption{The fitting curves of $\ln R_N$ vs. $\ln (1-\tilde{T})$ near the critical point for the reentrant phase transition. The red dots are numerical data and blue solid lines are obtained from the fitting formulas. (a) The coexistence SBH branch. (b) The coexistence LBH branch.}
\label{fig:fitline_14}
\end{figure}

The numerical results and the fitting results are shown in Fig. \ref{fig:fitline_14} with highly consistent with each other. The fitting coefficients of the small and large black holes are respectively $ \alpha_{LBH}$ =1.9536, $\alpha_{SBH}$ =2.0426. Taking into account the error of the calculation, we get the critical exponent $\alpha=2$. The dimensionless constant is
\begin{equation}\label{beta}
R_{N}(1-\tilde{T})^{2}=e^{(-2.5918-1.5253)/2}= -0.127636\simeq-\frac{1}{8},
\end{equation}
which is the same as that of the VdW-like phase transition.

\section{conclusion}\label{d}

In this paper, We have studied the phase transition and Ruppeiner geometry in different ranges of QED parameter in the extended phase space. For $a<0$ and $0 \leq a \leq \frac{32 Q^2}{7}$, we observed the small/large black hole phase transition and reentrant phase transition respectively. While for $\frac{32 Q^2}{7}<a$, there will be no the first-order phase transition. In these different ranges, we explored the black hole microstructure and different potential interactions are uncovered for the charged EH-AdS black hole.

In the first part, we investigated the thermodynamic properties of the black hole phase transition. Treating the cosmological constant and QED parameter as two new variables, we got the first law of the black hole thermodynamics and the Smarr formula hold. We also confirmed that they are consistent with each other in the extended phase space. Further, via the Hawking temperature, we obtained the equation of state. Employing it, the critical point is obtained. It is shown that for $a<0$, $0 \leq a \leq \frac{32 Q^2}{7}$, and $\frac{32 Q^2}{7}<a$, one, two, and zero critical points can be observed. According to it, we studied the phase transition in these parameter ranges.

For negative QED parameter $a$, we observed a characteristic swallow tail behavior of the Gibbs free energy below the critical point, which indicates there is the typical first-order black hole phase transition. The phase diagrams are also explicitly shown. When $0 \leq a \leq \frac{32 Q^2}{7}$, two critical points observed indicate a rich phase transition beyond the small/large black hole phase transition. For some certain values of the temperature or pressure, four black hole branches are found with two of them being unstable and the other two stable. From the behavior of the Gibbs free energy, these two stable branches form a reentrant phase transition. We then exhibited its phase diagrams, which is quite different from the small/large black hole phase transition of the VdW type. When the QED parameter has a large value such that $\frac{32 Q^2}{7}<a$, only one black hole branch is stable, and thus no black hole phase transition exists.

The second part is devoted to the study of microstructure by using the Ruppeiner geometry. Taking ($T$, $V$) as two fluctuation coordinates in thermodynamic phase space, we constructed the Ruppeiner geometry and calculated the corresponding normalized scalar curvature for the charged EH-AdS black holes. The scalar curvature behaves quite differently for the small/large black hole phase transition and reentrant phase transition, which may be due to the fact that they have different phase structures. For the small/large black hole phase transition, the normalized scalar curvature has two divergent points for each isothermal curve at most. While for the reentrant phase transition, an additional divergent point near small volume is observed.

Via the empirical observation of the Ruppeiner geometry, we found that for the black hole with negative QED parameter, the repulsive interaction among its microstructure will dominate for the small black hole with high temperature, while the attractive interaction dominates for other black holes. This result is similar to that of the charged AdS black hole without the QED parameter. For the charged EH-AdS black hole with small QED parameter, another region of positive scalar curvature emerges. It is below while not surrounded by the coexistence curve of the first-order phase transition. So for this case, different from the small/large black hole phase transition, the equation of state is applicable. Therefore, the low temperature black hole could have dominated repulsive interaction.

Furthermore, the behavior of the scalar curvature along the first-order coexistence curve of small and large black holes is carefully analyzed. For the case $a$=-1.5 and $Q$=1, the scalar curvature for both the coexistence small and large black holes decreases and goes to negative infinity at the critical temperature. However for $a$=1 and $Q$=1, a reentrant phase transition is present. Except for the divergent point near the critical point, the scalar curvature of the coexistence small black hole also diverges near $ \tilde{T}$=0.48, which is mainly because that the starting point of the coexistence small black hole is on the spinodal curve. This is also one of the novel features for the reentrant phase transition.

In particular, through the numerical calculation, we observed a critical exponent 2 and a dimensionless constant $-\frac{1}{8}$ near the critical point for the scalar curvature. This result is the same as that of other black holes and VdW fluid suggesting a result of the mean field theory.

Our work gives a complete study of the phase transition and microstructure for the charged EH-AdS black hole. The effects of the QED parameter on the black hole thermodynamics are examined in detail. Via the geometric method, the interaction among the microstructures is uncovered. These results help us peek into the nature of the black hole with a QED correction from the viewpoint of thermodynamics.

\section*{Acknowledgements}
This work was supported by the National Natural Science Foundation of China (Grants No. 12075103, No. 11675064, and No. 12047501).

\end{document}